\documentclass[
reprint,            
superscriptaddress, 
amsmath,            
amssymb,            
aps,                
prd,                 
floatfix,           
nofootinbib,        
]{revtex4-1}

\usepackage{tensor}     
\usepackage{graphicx}   
\usepackage[
colorlinks=true,        
citecolor=blue,         
linkcolor=blue,         
urlcolor=blue           
]{hyperref}             
\usepackage{bm}         
\usepackage{xcolor}     
\newcommand{\newc}{\newcommand*}
\newc{\figurewidth}{3.2in}
\newc{\xbar}{\bar{x}}
\newc{\rhoeq}{\rho_{\rm{eq}}}
\newc{\zeq}{z_{\rm{eq}}}
\newc{\la}{\lambda}
\newc{\tla}{\tilde{\la}}
\newc{\dt}{\delta}
\newc{\Dt}{\Delta}
\newc{\vj}{\vec{j}}
\newc{\vl}{\vec{l}}
\newc{\hx}{\hat{x}}
\newc{\hy}{\hat{y}}
\newc{\bj}{\bm{j}}
\newc{\mJ}{\mathcal{J}}
\newc{\mP}{\mathcal{P}}
\newc{\ga}{\gamma}
\newc{\Msun}{M_\odot}
\newc{\app}{\approx}
\newc{\av}[1]{\langle #1 \rangle}
\newc{\eq}[1]{Eq.~\eqref{#1}}
\newc{\al}{\alpha}
\newc{\Xstar}{X_{\ast}}
\newc{\seq}{\sigma_{\rm{eq}}}
\newc{\fpbh}{f_{\rm{pbh}}}
\newc{\RR}{{\cal R}}

\def\p{\partial}

\def\({\left(}
\def\){\right)}
\def\[{\left[}
\def\]{\right]}

\def\e{\begin{equation}}
\def\q{\end{equation}}
\def\m{\begin{eqnarray}}
\def\n{\end{eqnarray}}

\newc{\red}[1]{\textcolor{red}{#1}}
\newc{\yellow}[1]{\textcolor{yellow}{#1}}
\newc{\green}[1]{\textcolor{green}{#1}}
\newc{\blue}[1]{\textcolor{blue}{#1}}

\begin{document}

\title{
Merger rate distribution of primordial black hole binaries with electric charges}
\author{Lang Liu}
\email{liulang@itp.ac.cn}
\affiliation{CAS Key Laboratory of Theoretical Physics,
Institute of Theoretical Physics, Chinese Academy of Sciences,
Beijing 100190, China}
\affiliation{School of Physical Sciences,
University of Chinese Academy of Sciences,
No. 19A Yuquan Road, Beijing 100049, China}
\author{Zong-Kuan Guo}
\email{guozk@itp.ac.cn}
\affiliation{CAS Key Laboratory of Theoretical Physics,
Institute of Theoretical Physics, Chinese Academy of Sciences,
Beijing 100190, China}
\affiliation{School of Physical Sciences,
University of Chinese Academy of Sciences,
No. 19A Yuquan Road, Beijing 100049, China}
\author{Rong-Gen Cai}
\email{cairg@itp.ac.cn}
\affiliation{CAS Key Laboratory of Theoretical Physics,
Institute of Theoretical Physics, Chinese Academy of Sciences,
Beijing 100190, China}
\affiliation{School of Physical Sciences,
University of Chinese Academy of Sciences,
No. 19A Yuquan Road, Beijing 100049, China}
\author{Sang Pyo Kim}
\email{sangkim@kunsan.ac.kr}
\affiliation{CAS Key Laboratory of Theoretical Physics,
Institute of Theoretical Physics, Chinese Academy of Sciences,
Beijing 100190, China}
\affiliation{Department of Physics, Kunsan National University, Kunsan 54150, Korea}
\date{\today}
\begin{abstract}
We consider gravitational radiation and electromagnetic radiation from point mass binary with electric charges in a Keplerian orbit, and calculate the merger rate distribution of primordial black hole binaries with charges and a general mass function by taking into account gravitational torque and electromagnetic torque by the nearest primordial black hole. We apply the formalism to the extremal charged case and find that $\alpha=-(m_i+m_j)^2\partial^2 \ln {\cal R}(m_i,m_j)/\partial m_i \partial m_j=12/11$, which is independent of the mass function.
\end{abstract}

\pacs{???}

\maketitle
\section{Introduction}\label{intro}
Primordial black holes (PBHs)~\cite{Hawking:1971ei,Carr:1974nx,Carr:1975qj} are those black holes which are  formed in the very early Universe. Some mechanisms have been proposed to produce PBHs, such as large curvature perturbations generated during inflation~\cite{Stewart:1996ey,Stewart:1997wg,Leach:2000ea,Cheng:2018yyr,Gao:2018pvq}, domain walls \cite{Deng:2016vzb,Liu:2019lul}, bubble collisions~\cite{Kodama:1982sf,Hawking:1982ga,Lewicki:2019gmv}, preheating instability~\cite{Martin:2019nuw}, sound speed resonance~\cite{Cai:2018tuh} and parametric amplication of curvature perturbations~\cite{Cai:2019bmk}. Since LIGO  detected black hole binary mergers, PBHs,  as a promising candidate for dark matter (DM), have recently attracted much attention~\cite{Carr:2009jm,Cai:2018rqf,Belotsky:2018wph,Carr:2016drx,Sasaki:2018dmp,Cai:2018dig,Kannike:2017bxn,Kuhnel:2019xes,Gow:2019pok,Chen:2018rzo,Saito:2008jc,Wang:2019kaf,Chen:2019irf,Laha:2019ssq,Cai:2019elf,Chen:2019xse,Dasgupta:2019cae}. It is believed that the gravitational wave (GW) events observed by the LIGO detectors~\cite{Abbott:2016blz} could be explained by the coalescence of PBH binaries.\footnote{LIGO black holes also can be explained by stellar-origin black holes \cite{Belczynski:2016obo,Belczynski:2017gds}.} By calculating the late-time merger rate of PBHs which formed binaries in the late Universe, Refs \cite{Bird:2016dcv,Clesse:2016vqa} claim that the PBH merger rate could match the merger rate detected by LIGO if PBHs could account for all of the DM. In fact, there are two kinds of mechanisms proposed  for PBH binary formation. One is that PBH binary formed in the late Universe \cite{Bird:2016dcv,Clesse:2016vqa,Nishikawa:2017chy} while the other is that PBH binary formed in the early Universe \cite{Nakamura:1997sm,Ioka:1998nz,Sasaki:2016jop,Raidal:2017mfl,Kocsis:2017yty,Ali-Haimoud:2017rtz,Chen:2018czv,Ballesteros:2018swv,Raidal:2018bbj,Liu:2018ess,Liu:2019rnx,Young:2019gfc,Vaskonen:2019jpv,Garriga:2019vqu}, that is expected to make the dominant contribution to the PBH merger rate today. 

The merge rate of PBH binaries with monochromatic mass function is estimated through the three-body interaction~\cite{Nakamura:1997sm,Ioka:1998nz,Sasaki:2016jop}. Later, the merger rate of PBH binaries is improved in \cite{Ali-Haimoud:2017rtz} by taking into account the torques exerted by all PBHs, but it is also assumed that all PBHs have the same mass. The mechanism has recently been developed for a general mass function by taking into account the torques from the all PBHs~\cite{Chen:2018czv,Raidal:2018bbj,Liu:2018ess}. A formalism to estimate the effect of merger history of PBHs on merger rate distribution has been developed in \cite{Liu:2019rnx}. Those works consider the merger rate distribution of PBHs binaries  by assuming that PBHs are Schwarzschild black holes. However, in general case, PBHs have spin and charges. In this paper, we study the emission of gravitational and electromagnetic waves from binaries of charged black holes and find the merger rate density for these binaries with arbitrary ratio of charge to mass and wide range of masses. A simple ratio of characterizing parameter  $\alpha =-(m_i+m_j)^2\partial^2 \ln {\cal R}(m_i,m_j)/\partial m_i \partial m_j$ is first derived in Ref. \cite{Kocsis:2017yty} to distinguish PBH and stellar-origin BHs.  For astrophysical mechanisms leading to black hole mergers are generally expected to yield different values. Reference \cite{O_Leary_2016}  shows that the probability of merger is proportional to $(m_i+m_j)^4$ for binary black hole mergers in dense star clusters, which implies $\alpha=4$. For the uncharged PBH case, $\alpha = 36/37$ \cite{Raidal:2018bbj,Liu:2018ess}. In this paper, we find $\alpha=12/11$ which is independent of the mass function for binaries of extremal charged black holes in contrast to $\alpha = 36/37$ for uncharged PBH binaries.


Charged black holes have attracted much attention not only in theoretical study of Hawking radiation and Schwinger effect but also in recent observations of GWs. A non-extremal charged black hole emits all species of particles, neutral or charged, according to the Bose-Einstein or Fermi-Dirac distribution with the Hawking temperature (\cite{Page:2004xp} for a review). The Hawking temperature vanishes for extremal charged black holes, which may literally cease the evaporation. The Schwinger mechanism, however, triggers pair creation of charged particles from extremal black holes~\cite{Chen:2017mnm}. The leading Boltzmann factor is given by the effective temperature for accelerated charges in the electric field on the horizon \cite{Cai:2014qba}, whose near-horizon geometry has a factor of ${\rm AdS}_2$ space.

When the horizon size of a PBH is smaller than the Compton wavelength or classical radius of a charged particle, the PBH cannot emit  particles and may be a candidate for dark matter~\cite{Aharonov:1987tp}. For (near-) extremal charged black holes, this is equivalent to the Breitenlohler-Friedmann (BF) bound since the  ${\rm AdS}_2$ geometry near the horizon gives the bound $|R_{\rm AdS}|/2 \geq (qE_H/\bar{m})^2$ against the Schwinger mechanism~\cite{Pioline:2005pf,Cai:2014qba}, which in turn gives the BH size $2|q| \geq R_{H}$ for the charge $q$ ~\cite{Chen:2016caa}  and the mass bound $M \leq \frac{m_P}{2 \sqrt{q^2 - (\frac{m}{m_P})^2}}$. These extremal PBHs have small masses and may also be a candidate for dark matter. On the other hand, in the early universe and beyond the standard model, a dark quantum electrodynamics with heavy dark electrons and massless dark photons, which couple to electrons and photons of the standard model at renormalization level, suppresses the Schwinger effect and allows the extremal PBHs whose life time is longer than the age of the universe~\cite{Bai:2019zcd}. These dark electric charges  have a hidden U(1) symmetry and are formally described by the same Maxwell theory.  In this paper we assume such scenarios for extremal PBHs.

The paper is organized as follows. In the next section, we calculate gravitational radiation and electromagnetic radiation from point masses with charges in a Keplerian orbit. In Sec.~\ref{Formalism},  we derive the merger rate distribution of PBH binaries with charges and a general mass function by taking into account gravitational torque and electromagnetic torque by the nearest PBH.
In Sec.~\ref{Cases}, we consider a specific cases of extremal charged PBH binaries, we find that $\alpha=12/11$, which is independent of the mass function.
The last section is devoted to conclusions and discussions.

In this paper, we choose units of $c =\epsilon_0 = \mu_0=1$. Whenever relevant, we adopt the values of cosmological parameters from the \emph{Planck 2018} results~\cite{Aghanim:2018eyx} and the scale factor $s(t)$ is normalized to be unity at the matter-radiation equality.

\section{Electromagnetic radiation and gravitational radiation}\label{GW and EW}
\begin{figure}[htbp!]
\centering
\includegraphics[width = 0.6\textwidth]{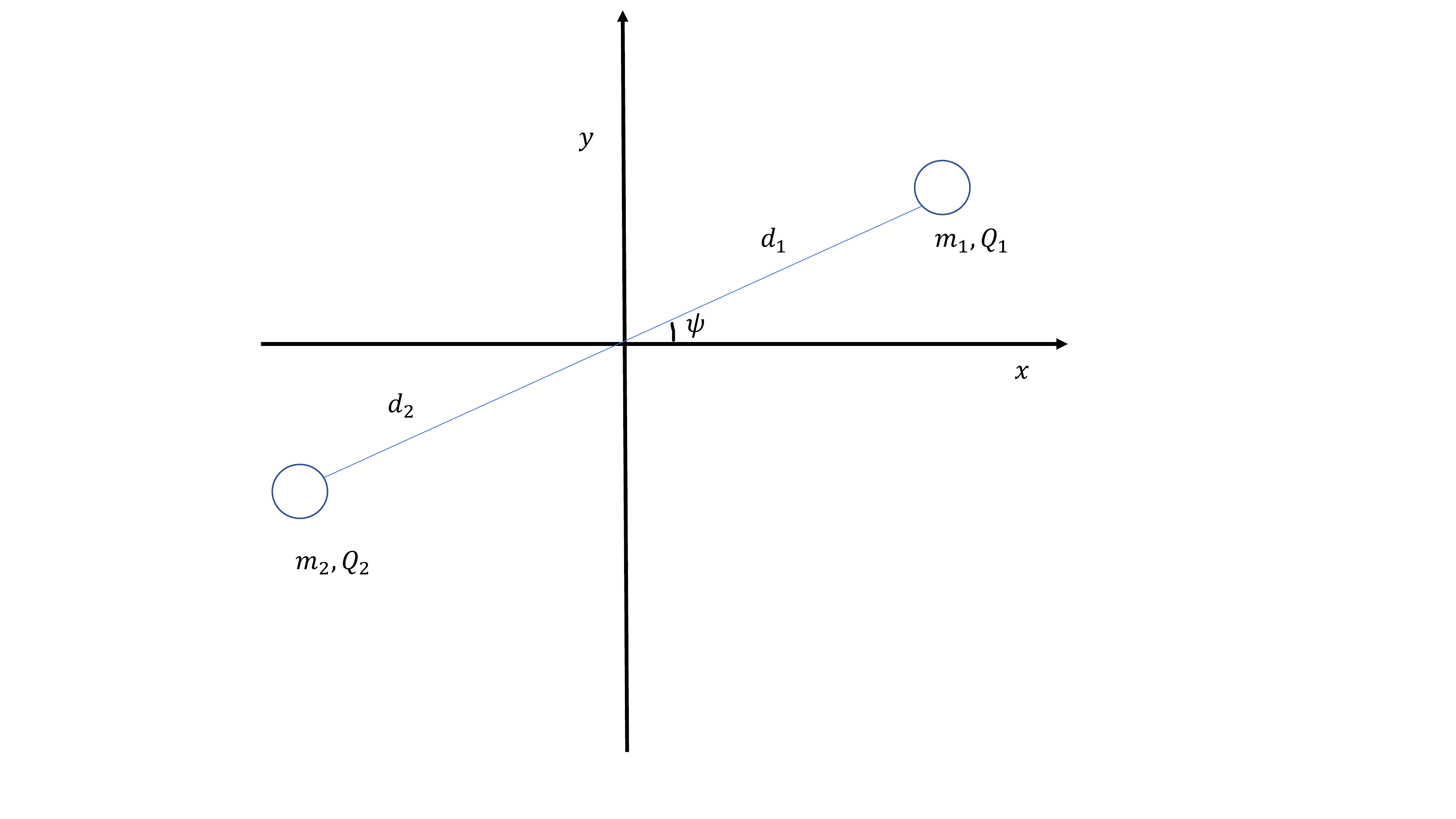}
\caption{\label{fig:1}
A schematic picture of point masses with charges in a Keplerian orbit. }
\end{figure}

The point masses $m_1$ with charge $Q_1$ and $m_2$ with charge $Q_2$ have coordinates ($d_1\cos{\psi}$, $d_1 \sin{\psi}$) and ($-d_2 \cos{\psi}$, $-d_2\sin{\psi}$) in the $x$-$y$ plane, as shown in Fig.~\ref{fig:1}. Choosing the origin to be the center of mass, we have
\m
\label{d1d2}
d_{1}=\left(\frac{m_{2}}{m_{1}+m_{2}}\right) d, \quad d_{2}=\left(\frac{m_{1}}{m_{1}+m_{2}}\right) d,
\n
where $d=d_1+d_2$  is the distance between the two point masses. The total energy is given by
\m
\label{E1}
E=-\frac{Gm_1m_2}{2a}+\frac{1}{4\pi}\frac{Q_1Q_2}{2a}=-\frac{Gm_1m_2}{2a}(1-\lambda ),
\n
where $a$ is the semi-major axis and
\m
\lambda=\frac{1}{4\pi}\frac{Q_1Q_2}{Gm_1m_2}.
\n
Because the point masses make up a bound system, we have $\la<1$.
For the Kepler motion, the orbit equation, angular velocity and angular momentum are given by
\m
d=\frac{a\left(1-e^{2}\right)}{1+e \cos \psi},
\n
\m
\label{psi}
\dot{\psi}=\frac{\left[G\left(m_{1}+m_{2}\right) a\left(1-e^{2}\right) (1-\la)\right]^{1 / 2}}{d^{2}},
\n
\m
\label{TL}
L=\frac{\sqrt{a} \sqrt{1-e^2} \sqrt{G} \sqrt{1-\lambda } m_1 m_2}{\sqrt{m_1+m_2}},
\n
where $e$ is the eccentricity. Firstly, we compute the total power radiated in electromagnetic waves. In our reference frame where the orbit is in the $x$-$y$ plane, the electric dipole is given by
\m
\label{PQ}
\bm{p}&\equiv& Q_1 \bm{x_1}+Q_2 \bm{x_2}
\notag \\
&=&\frac{m_2Q_1-m_1Q_2}{m_1+m_2} d \cos \psi \hat{\bm{x}}+\frac{m_2Q_1-m_1Q_2}{m_1+m_2} d \sin \psi \hat{\bm{y}},
\notag \\
\n
where $\hat{\bm{x}}$ is the unit vector along $\bm{x}$ and $\hat{\bm{y}}$ is the unit vector along $\bm{y}$. The Lagrangian density of the electromagnetic field is
\m
\mathcal{L}_{EM}=-\frac{1}{4} F_{\mu \nu} F^{\mu \nu}=\frac{1}{2}\left(\mathbf{E}^{2}-\mathbf{B}^{2}\right).
\n
The electric field $E$ and magnetic field $B$ at $\bm{r}$ $(r\gg d)$ are
\m
\label{E}
\mathbf{E}(\mathbf{r}, t) \cong \frac{1}{4 \pi r}[\hat{\mathbf{r}} \times(\hat{\mathbf{r}} \times \ddot{\mathbf{p}})],
\n
\m
\label{B}
\mathbf{B}(\mathbf{r}, t) \cong-\frac{1}{4 \pi r }[\hat{\mathbf{r}} \times \ddot{\mathbf{p}}],
\n
where $\hat{\bm{r}}$ is the unit vector along $\bm{r}$ and $t^{\prime}=t-r$. Because of emitting electromagnetic radiation, the system loses energy and angular momentum.
From
\m
\partial_{\mu} T_{EM}^{\mu \nu}=0,  \quad E_{EM}=\int_V d x^3 T_{EM}^{00},
\n
the rate of energy emission due to electromagnetic radiation is
\m
\frac{dE_{EM}}{dt}=-\int_V d x^3 \p_i T_{EM}^{0i}=-\int_S d \vec{A} \cdot \vec{n}_i T_{EM}^{0i}=-\frac{\ddot{p}^2}{6\pi},
\notag \\
\n
where $T_{EM}^{\mu \nu}$ is the energy-momentum tensor of the electromagnetic field. By using Eq.~\eqref{PQ}, we have the average energy loss over an orbital period $T$ due to electromagnetic radiation
\m
\left\langle\frac{dE_{EM}}{dt}\right\rangle &\equiv& \frac{1}{T} \int_{0}^{T} d t \frac{dE_{EM}}{dt}
\notag \\
&=& \frac{\left(e^2+2\right) G^2 (1-\lambda)^2 (m_2 Q_1-m_1 Q_2)^2}{12 \pi  a^4 \left(1-e^2\right)^{5/2}},
\notag \\
\n
where
\m
\label{T}
T=\int_{0}^{2\pi} d \psi \dot{\psi}^{-1} = \frac{2 \pi  a^2}{\sqrt{a G (1-\lambda ) (m_1+m_2)}}.
\n

The angular momentum of the electromagnetic field  along the $i$ axis is then given by $J_{EM}^{i}=(1 / 2) \epsilon^{i j k} J_{EM}^{j k}$. From the Noether’s theorem, we have
\m
J_{EM}^{j k}&=&\int d^{3} x j_{EM}^{0(j k)}
\notag \\
&=&\int d^{3} x\left[\frac{\partial \mathcal{L}_{EM}}{\partial\left(\partial_{0} A_{i}\right)}\left(a^{\nu(j k)} \partial_{\nu} A_{i}-F_{i}^{(j k)}\right)-a^{0(j k)} \mathcal{L}_{EM}\right],
\notag \\
\n
where
\m
a_{(\rho \sigma)}^{\mu}=\delta_{\rho}^{\mu} x_{\sigma}-\delta_{\sigma}^{\mu} x_{\rho}, \quad F^{i(j k)}=\delta^{i j} A^{k}-\delta^{i k} A^{j}.
\n
After a straightforward computation, we obtain
\m
J_{EM}^{i}=\int d^{3} x\left[-\epsilon^{i k l}\left(\partial_{0} A_{j}\right) x^{k} \partial^{l} A_{j}+\epsilon^{i k l} A_{k} \partial_{0} A_{l}\right],
\notag \\
\n
where the first term is the orbital angular momentum and the second term is the spin part. The density of the angular momentum of the electromagnetic field  is given by
\m
j_{EM}^{i}=-\epsilon^{i k l}\left(\partial_{0} A_{j}\right) x^{k} \partial^{l} A_{j}+\epsilon^{i k l} A_{k} \partial_{0} A_{l}.
\n
Let us consider electromagnetic waves propagating outward from the two point masses. At time $t$ we consider a portion of the wave front covering a solid angle $d \Omega$ at radial distance $r$ from our source, and then at time $t+d t$ , this portion of the wave front has swept the volume $d^{3} x=r^{2} d r d \Omega=r^{2}(d t) d \Omega $. Since the angular momentum of electromagnetic waves per unit volume is $j_{EM}^{i}$ , the angular momentum carried away by electromagnetic waves is given by
\m
d J_{EM}^{i}=r^{2} d t d \Omega j_{EM}^{i} .
\n
Therefore the rate of angular momentum emission due to electromagnetic waves is obtained by
\m
\frac{d J_{EM}^{i}}{d t}=- \int r^{2} d \Omega (-\epsilon^{i k l}\left(\partial_{0} A_{j}\right) x^{k} \partial^{l} A_{j}+\epsilon^{i k l} A_{k} \partial_{0} A_{l}).
\notag \\
\n
According to Appendix\ref{AngularMomentum}, we obtain
\m
\frac{d J_{EM}^{i}}{d t}= -\frac{\epsilon^{i k l}}{6\pi}\dot{p}_k \ddot{p}_l.
\n
For the orbit in the $x$-$y$ plane, we have $L_z=L, L_x=L_y=0$. Using Eqs.~\eqref{PQ} and ~\eqref{psi}, one has
\m
\dot{p}_1=\frac{\sqrt{G} \sqrt{1-\lambda } \sin (\psi ) (m_2 Q_1-m_1 Q_2)}{\sqrt{a} \sqrt{1-e^2} \sqrt{m_1+m_2}},
\n
\m
\ddot{p}_1=-\frac{G (1-\lambda) \cos (\psi ) (e \cos (\psi )+1)^2 (m_2 Q_1-m_1 Q_2)}{a^2 \left(1-e^2\right)^2},
\notag \\
\n
\m
\dot{p}_2=\frac{\sqrt{G} \sqrt{1-\lambda} (e+\cos (\psi )) (m_1 Q_2-m_2 Q_1)}{\sqrt{a} \sqrt{1-e^2} \sqrt{m_1+m_2}},
\n
\m
\ddot{p}_2=-\frac{G (1-\lambda ) \sin (\psi ) (e \cos (\psi )+1)^2 (m_2 Q_1-m_1 Q_2)}{a^2 \left(1-e^2\right)^2}.
\notag \\
\n
The rate of angular momentum emission due to electromagnetic radiation is given by
\m
\frac{d J_{EM}}{d t}&=& -\frac{1}{6\pi} (\dot{p}_2 \ddot{p}_1-\dot{p}_1 \ddot{p}_2)
\notag \\
&=&-\frac{G^{3/2} (1-\lambda)^{3/2} (e \cos (\psi )+1)^3 (m_2 Q_1-m_1 Q_2)^2}{6 \pi  a^{5/2} \left(1-e^2\right)^{5/2} \sqrt{m_1+m_2}}.
\notag \\
\n
For the angular momentum loss due to electromagnetic radiation averaged one orbital period $T$, we have
\m
\left\langle\frac{dJ_{EM}}{dt}\right\rangle &\equiv& \frac{1}{T} \int_{0}^{T} d t \frac{dJ_{EM}}{dt}
\notag \\
&=&-\frac{ G^{3/2} (1-\lambda )^{3/2} (m_2 Q_1-m_1 Q_2)^2}{6 \pi  a^{5/2} \left(1-e^2\right) \sqrt{m_1+m_2}}.
\n
The electromagnetic field or gravitational field carries away a total angular momentum $J$, which is made of a spin contribution and of an orbital angular momentum contribution. This total angular momentum is drained from the total angular momentum of the source, which, for our binary system or any macroscopic source, is a purely orbital angular momentum. So, the loss rate of the angular momentum in the system due to electromagnetic radiation is given by
\m
\left\langle\frac{dL_{EM}}{dt}\right\rangle &=&\left\langle\frac{dJ_{EM}}{dt}\right\rangle
\notag \\
&=&-\frac{ G^{3/2} (1-\lambda )^{3/2} (m_2 Q_1-m_1 Q_2)^2}{6 \pi  a^{5/2} \left(1-e^2\right) \sqrt{m_1+m_2}}.
\n
Now, we begin to compute the total radiated power in GWs. In our reference frame where the orbit is in the $x$-$y$ plane, the second mass moment is given by a $2 \times 2$ matrix
\begin{equation}
\label{Mab}
M_{a b}=\mu d^{2}\left(\begin{array}{cc}{\cos ^{2} \psi} & {\sin \psi \cos \psi} \\ {\sin \psi \cos \psi} & {\sin ^{2} \psi}\end{array}\right)_{a b},
\end{equation}
where $\mu=m_1m_2/(m_1+m_2)$ is the reduced mass and subscripts $\(a, b=1,2\)$ are indices in the $x$-$y$ plane.
Following \cite{Peters:1963ux}, the radiated  power of GWs can be expressed in a rotation invariant form
\m
\label{P}
P(\psi)= \frac{2 G}{15 }\left[\bigl( \dddot{M}_{11} + \dddot{M}_{22} \bigr)^2 - 3 \bigl( \dddot{M}_{11} \dddot{M}_{22} -  \dddot{M}_{12}^{2} \bigr) \right],
\notag \\
\n
where the first term is the square of the trace and the second term is the determinant of the matrix of $\dddot{M}_{ij}$.
Using Eqs.~\eqref{psi} and \eqref{Mab}, one has the components of the matrix
\m
\label{M12}
\dddot{M}_{11}&=&\frac{G^{3/2} (1-\lambda)^{3/2} m_1 m_2 \sqrt{m_1+m_2} }{a^{5/2} \left(1-e^2\right)^{5/2}}
\notag \\
&\times& \sin (2 \psi ) (e \cos (\psi )+1)^2 (3 e \cos (\psi )+4) ,
\n
\m
\dddot{M}_{12}&=&-\frac{G^{3/2} (1-\lambda )^{3/2} m_1 m_2 \sqrt{m_1+m_2} (e \cos (\psi )+1)^2}{2 a^{5/2} \left(1-e^2\right)^{5/2}}
\notag \\
&\times&  (5 e \cos (\psi )+3 e \cos (3 \psi )+8 \cos (2 \psi )),
\n
\m
\dddot{M}_{22}&=&-\frac{G^{3/2} (1-\lambda )^{3/2} m_1 m_2 \sqrt{m_1+m_2} (e \cos (\psi )+1)^2}{a^{5/2} \left(1-e^2\right)^{5/2}}
\notag \\
&\times&  \sin (\psi ) (e (3 \cos (2 \psi )+5)+8 \cos (\psi )).
\n
So, we get
\m
P(\psi)&=& \frac{4 G^4 (1-\lambda )^3 m_1^2 m_2^2 (m_1+m_2)  (e \cos (\psi )+1)^4}{15 a^5 \left(1-e^2\right)^5}
\notag \\
&\times& \left(11 e^2 \cos (2 \psi )+13 e^2+48 e \cos (\psi )+24\right) .
\nonumber \\
\n
The energy of GWs is only well-defined by taking an average over several periods . In our case, a well-defined quantity is the average of $P(\psi)$ over one period $T$. So we can perform this time average to get the total radiated power
\m
\bar{P} &\equiv& \frac{1}{T} \int_{0}^{T} d t P(\psi)
\notag \\
&=&\frac{\left(37 e^4+292 e^2+96\right) G^4 (1-\lambda )^3 m_1^2 m_2^2 (m_1+m_2)}{15 a^5  \left(1-e^2\right)^{7/2}}.
\notag \\
\n
The average energy loss over an orbital period $T$ is given by
\m
&&\left\langle\frac{dE_{GW}}{dt}\right\rangle =-\bar{P}
\notag \\
&=&-\frac{\left(37 e^4+292 e^2+96\right) G^4 (1-\lambda )^3 m_1^2 m_2^2 (m_1+m_2)}{15 a^5  \left(1-e^2\right)^{7/2}}.
\notag \\
\n
Following \cite{Peters:1964zz}, the rate of angular momentum emission due to GWs is given by
\m
\label{dLgw}
\frac{d L_{GW}^{i}}{d t}=-\frac{2 G}{5 } \epsilon^{i k l}\left\langle\ddot{M}_{k a} \dddot{M}_{l a}\right\rangle.
\n
For the orbit in the $x$-$y$ plane,  we have $L_z=L, L_x=L_y=0$. So, we get
\m
\frac{d L_{GW}}{d t}&=&\frac{4 G}{5 }\left\langle\ddot{M}_{12}\left(\dddot{M}_{11}-\dddot{M}_{22}\right)\right\rangle.
\n
For the angular momentum loss  averaged over one orbital period $T$, we have
\m
\left\langle\frac{d L_{GW}}{d t}\right\rangle&=&\frac{1}{T} \int_{0}^{T} d t \frac{dL_{GW}}{dt}
\notag \\
&=&-\frac{4  \left(7 e^2+8\right) G^{7/2} (1-\lambda )^{5/2} m_1^2 m_2^2 \sqrt{m_1+m_2}}{5 a^{7/2}  \left(1-e^2\right)^2}.
\notag \\
\n
The total rate of energy and angular momentum emission due to electromagnetic radiation and gravitational radiation are given by
\m
\left\langle\frac{dE}{dt}\right\rangle=\left\langle\frac{dE_{EM}}{dt}\right\rangle+\left\langle\frac{dE_{GW}}{dt}\right\rangle,
\n
\m
\left\langle\frac{dL}{dt}\right\rangle=\left\langle\frac{dL_{EM}}{dt}\right\rangle+\left\langle\frac{dL_{GW}}{dt}\right\rangle.
\n
Whatever $e \simeq 0$ or $e\simeq 1$, we have
\m
\frac{\left\langle\frac{dE_{GW}}{dt}\right\rangle}{\left\langle\frac{dE_{EM}}{dt}\right\rangle}  &\approx& \frac{\left\langle\frac{dL_{GW}}{dt}\right\rangle}{\left\langle\frac{dL_{EM}}{dt}\right\rangle}
\notag \\
 &\sim& \frac{24 \pi  \left(7 e^2+8\right) G^2 (1-\lambda) m_1^2 m_2^2 (m_1+m_2)}{5 a \left(1-e^2\right)  (m_2 Q_1-m_1 Q_2)^2}.
 \notag \\
\n
The system spends most of the decay time in a state for which $a \approx a_0$ where $a_0$ and $e_0$ are initial conditions of the semi-major axis and the eccentricity. For a given $a_0$ and $e_0$, the total rate of energy and angular momentum emission is dominated by gravitational radiation or electromagnetic radiation which depends on $m_1,m_2,Q_1$ and $Q_2$.
If gravitational radiation is dominated, the coalescence time
\m
\tau_{GW} \approx \begin{cases}
\frac{5 a_0^4 }{256 G^3 (1-\lambda )^2 m_1 m_2 (m_1+m_2)}, ~~~~{\rm for}~e_0 \simeq 0,\\
   \frac{3 a_0^4 \left(1-e_0^2\right)^{7/2} }{85 G^3 (1-\lambda )^2 m_1 m_2 (m_1+m_2)},~~~~{\rm for}~e_0 \simeq 1.
  \end{cases}
\n
Similarly, if electromagnetic radiation is dominated, the coalescence time
\m
\tau_{EM} \approx \begin{cases}
\frac{\pi  a_0^3  m_1 m_2}{G (1-\lambda ) (m_2 Q_1-m_1 Q_2)^2}, ~~~~{\rm for}~e_0 \simeq 0,\\
    \frac{4 \pi  a_0^3 \left(1-e_0^2\right)^{5/2} m_1 m_2}{G (1-\lambda ) (m_2 Q_1-m_1 Q_2)^2},~~~~{\rm for}~e_0 \simeq 1.
  \end{cases}
\n
The coalescence time for two point masses with charges can be approximated as
\m
\tau\cong  {\rm Min}(\tau_{GW},\tau_{EM}).
\n

\section{Merger rate distribution of primordial black hole binaries with charges}
\label{Formalism}
Let us consider the condition two nearest PBHs with masses $m_i$, $m_j$ and charges $Q_i$, $Q_j$ decouple from the expanding Universe, assuming negligible initial peculiar velocities in what follows. The total energy and angular momentum of the bound system are
\m
\label{E1}
E=-\frac{Gm_im_j}{2a}+\frac{1}{4\pi}\frac{Q_iQ_j}{2a}=-\frac{Gm_im_j}{2a}(1-\lambda ),
\n
\m
L=\frac{\sqrt{a} \sqrt{1-e^2} \sqrt{G} \sqrt{1-\lambda } m_i m_j}{\sqrt{m_i+m_j}}.
\n
Considering the gravitational force, electromagnetic force and the expansion of the Universe,
the equation of motion for their proper distance $r$ in Newtonian approximation is given by
\m
\label{eom1}
  \ddot{r} - \( \dot{H} + H^2 \) r + \frac{m_b}{r^2} \frac{r}{|r|}(1-\la) = 0,
\n
where  $m_b=m_i+m_j$ is total mass of the PBH binary and the dot denotes the differentiation with respect to the proper time. By defining $\chi \equiv r/x$, we can rewrite Eq.~\eqref{eom1} as
\m
  \chi'' + \frac{s h' + h}{s^2 h} \(s \chi' - \chi \) + \frac{1}{\tilde {\la}}
    \frac{1}{\(sh\)^2} \frac{1}{\chi^2} \frac{\chi}{|\chi|} = 0,
    \label{chi}
\n
where primes denote differentiation with respect to the scale factor $s$, $h(s)\equiv H(s)/\({8\pi\over 3}\rhoeq\)^{1/2}=\sqrt{s^{-3}+s^{-4}}$, $\rho_{\rm{eq}}$ is the energy density of the Universe at the matter-radiation equality, and $x$ is the comoving separation between these two nearest PBHs. Here, the dimensionless parameter $\tilde{\la}$ is given by
\m
\label{la}
  \tilde{\la} = \frac{8 \pi \rhoeq x^3}{3 m_b(1-\la)}.
\n
 The solution of Eq.~\eqref{chi} derived in \cite{Ali-Haimoud:2017rtz} implies the semi-major axis $a$ of the formed binary is given by
\m
  a \approx 0.1 \tilde{\la} x .
      \label{axis}
\n

\begin{figure}[htbp!]
\centering
\includegraphics[width = 0.48\textwidth]{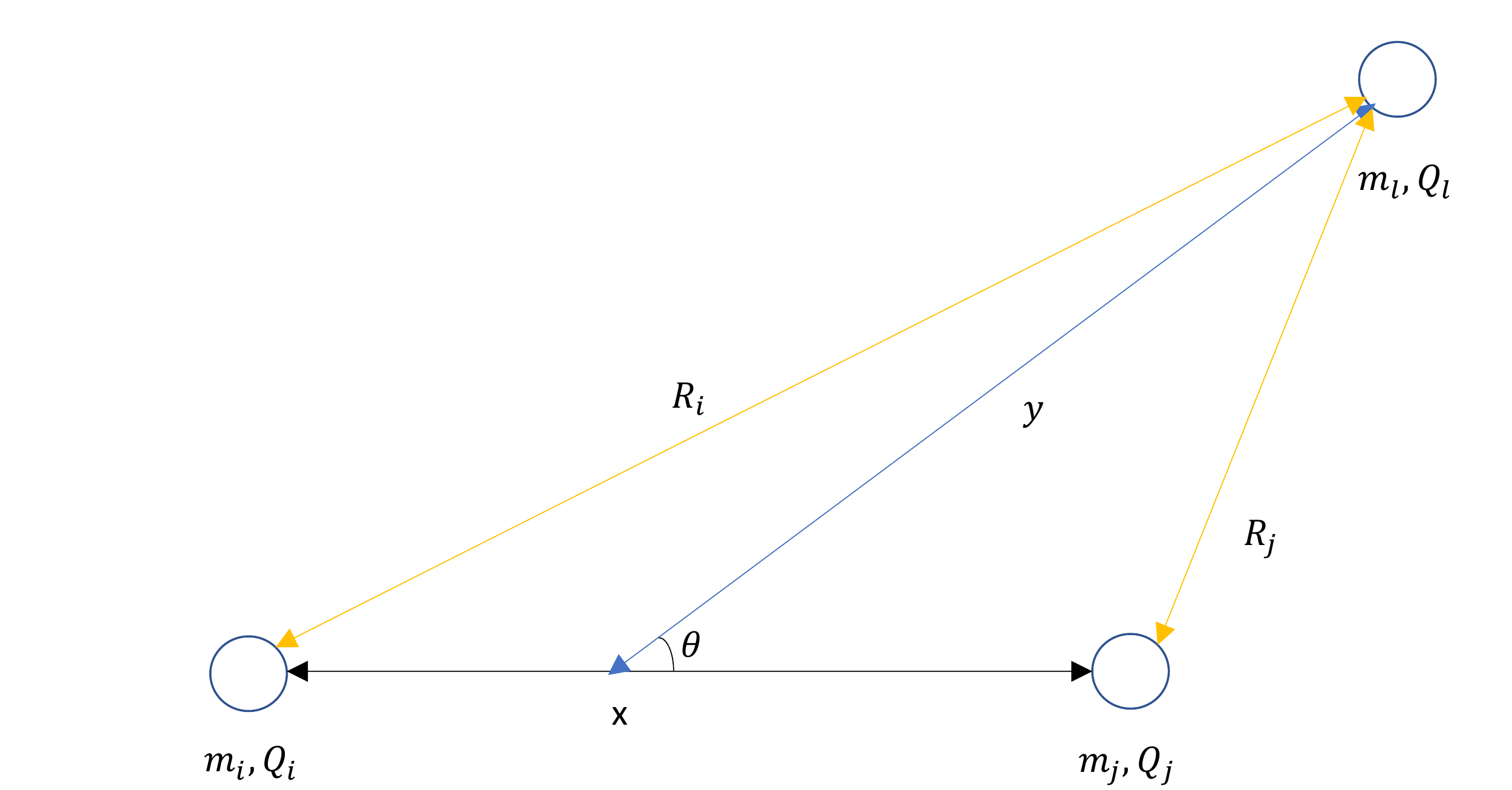}
\caption{\label{fig:2}
A schematic picture of the total exerted torques. }
\end{figure}

Then the two PBHs which could form a bound system come closer and closer, the surrounding PBHs, especially the nearest PBH, will exert torques on the PBH binary. The tidal force from other PBHs will provide an angular momentum to prevent this system from direct coalescence and form a highly eccentric binary.  The angular momentum $L$ of the binary is estimated by multiplying the exerted total torques $\ell$ from the nearest PBH with mass $m_l$ and charge $Q_l$ by the free-fall time
\m
L \approx |t_{ff} \times \ell|,
\n
where the free-fall time is given by
\m
t_{ff} \approx\frac{\pi}{2}\frac{x^{\frac{3}{2}}}{\sqrt{2Gm_b(1-\la)}}.
\n
The exerted total torque $\ell$ is made of the torque from the gravitational force $\ell_{GW}$ and the torque from electromagnetic force $\ell_{EM}$. As illustrated in Fig.~\ref{fig:2}, $y \gg x$ is the comoving distance from the third PBH to the PBH binary and $\theta$ is the angle between $\bm{x}$ and $\bm{y}$. Here, we introduce a dimensionless charge $k$
\m
k =\frac{Q}{\sqrt{4\pi G} m },
\n
so, $k_i$ represents $\frac{Q_i}{\sqrt{4\pi G} m_i }$. The torque  from the gravitational force $\ell_{GW}$ is given by
\m
\label{ellGW}
\ell_{GW} =\ell^i_{GW}+\ell^j_{GW}=-\frac{Gm_l m_i m_j x y \sin \theta}{R_i^3 m_b}+\frac{Gm_l m_i m_j x y \sin \theta}{R_j^3 m_b}.
\notag \\
\n
Using
\m
R_i=(y^2 \sin^2 \theta+(y\cos \theta +\frac{m_j}{m_b} x)^2)^{\frac{1}{2}} \approx y(1+\frac{m_j x \cos \theta}{m_by} ),
\notag \\
\n
\m
R_j=(y^2 \sin^2 \theta+(y\cos \theta -\frac{m_i}{m_b} x))^{\frac{1}{2}} \approx y(1-\frac{m_i x \cos \theta}{m_by} ),
\notag \\
\n
we can rewrite Eq.~\eqref{ellGW} as
\m
\ell_{GW} \approx \frac{3Gm_l m_i m_j x^2 \sin \theta \cos \theta}{m_b y^3}.
\n
Similarly, the torque from the electromagnetic force $\ell_{EM}$ is given by
\m
\ell_{EM} &\approx& -\frac{Gm_l m_i m_j x \sin \theta }{m_b y^2}
\notag \\
&\times& (k_j k_l-k_ik_l+\frac{3x\cos \theta }{m_b y}(m_i k_j k_l+m_j k_i k_l)).
\n
The total torque $\ell$ is given by
\m
\ell=-\frac{Gm_l m_i m_j x \sin \theta }{m_b y^2} F,
\n
where
\m
F=k_j k_l-k_ik_l+\frac{3x\cos \theta }{m_b y}(m_i k_j k_l+m_j k_i k_l-m_b).
\notag \\
\n
Now, we introduce a dimensionless angular momentum
\m
j \equiv  \sqrt{1-e^2}.
\n
By solving
\m
L \approx |t_{ff} \times \ell|,
\n
\m
L=\frac{\sqrt{x} j \sqrt{G} \sqrt{1-\lambda } m_i m_j}{\sqrt{m_i+m_j}},
\n
we can get
\m
j \approx \frac{x^2m_l \sin \theta}{y^2 m_b (1-\la)} |F|.
\n
For Schwarzschild black holes where $k_i=k_j=k_l=0$, we have
\m
j=3 \frac{m_l x^3 \sin \theta \cos \theta}{m_b y^3},
\n
which is consistent with the result in \cite{Chen:2018czv}. The coalescence time of PBH binaries derived in Sec.~\ref{GW and EW} can be estimated as
\m
&&\tau={\rm Min}
\notag \\
&&(\frac{3 a^4 j^{7} }{85 G^3 (1-\lambda )^2 m_i m_j (m_i+m_j)}, \frac{  a^3  j^{5} }{G^2 (1-\lambda )m_i m_j (k_i-k_j)^2})
\notag \\
\n
The probability distribution function of PBH masses and charges $P(m, k)$ is normalized to be
\m
\int_{-1}^{+1}\int_0^\infty dmdk P(m, k) =1.
\n
The abundance of PBHs with charges in the mass interval $(m, m+dm)$ is
\m
f P(m)dm, 
\n
where
\m
P(m)\equiv \int_{-1}^{+1} dkP(m, k).
\n

The fraction of PBHs in DM, $f_{\rm{pbh}}$, is related to the total abundance of PBHs in non-relativistic matter$f$ by $f_{\rm{pbh}}\equiv \Omega_{\rm{pbh}}/\Omega_{\rm{dm}} \approx f/0.85$.
The average number density of PBHs in mass interval $(m, m+dm)$ at the matter-radiation equality is given by
\m
n(m)dm= \frac{fP(m)dm \rho_{\rm{eq}}}{m},
\n
while the comoving total average number density of PBHs, $n_{T}$, is defined by
\m
n_T\equiv f\rho_{\rm{eq}}\int_0^\infty dm {P(m)\over m}.
\n
For simplicity, we could define $m_{\rm pbh}$ as
\m
 \label{mpbh}
\frac{1}{m_{\rm{pbh}}}\equiv \int_0^\infty dm {P(m)\over m} .
\n
So, ${n\(m\)}/{n_{T}}=P\(m\) {m_{\rm{pbh}}}/{m}$ is the fraction of the average number density of PBHs with mass $m$ in the total average number density of PBHs.

To calculate the merger rate of PBH binaries, we have to know the spatial distribution of PBHs.
Assuming that the spatial distribution of PBHs is random one,
for the comoving distances, $x$ and $y$, in the intervals $(x, x+dx)$ and $(y, y+dy)$, PBH masses, $m_i$, $m_j$ and $m_l$, in the intervals $(m_i, m_i+dm_i)$, $(m_j, m_j+dm_j)$ and $(m_l, m_l+dm_l)$, PBH charges, $k_i$, $k_j$ and $k_l$, in the intervals $(k_i, k_i+dk_i)$, $(k_j, k_j+dk_j)$ and $(k_l, k_l+dk_l)$, and the angle $\theta$, in the intervals $(\theta, \theta+d\theta)$, the probability is given by
\m \label{distribution 1}
&d{\mathcal{P}}&=P\(m_i, k_i\)dm_idk_i P\( m_j, k_j\)dm_jdk_j  P\(m_l, k_l\)dm_ldk_l
\nonumber \\
&\times& \frac{m_{\rm{pbh}}^3}{m_i m_j m_l} 4\pi x^2 n_{T}dx 2\pi y^2 \sin(\theta) n_{T}dy d\theta e^{- {4\pi \over 3}y^3 n_{T}} \Theta(y-x),
\notag \\
\n
The fraction of PBHs that have merged before the time $t$ is given by
\m
&&{G}(t,m_i,m_j,m_l,k_i,k_j,k_l)
\nonumber \\
&=&\int dxdyd\theta \frac{d \mathcal{P}}{dx dy dm_i dm_j dm_l dk_i dk_j dk_l d\theta} \Theta (t-\tau) .
\nonumber \\
\n
The merger rate density $\RR(t,m_i,m_j,m_l,k_i,k_j,k_l)$ is given by
\m
 \label{Rbar 1}
&\RR&(t,m_i,m_j,m_l,k_i,k_j,k_l)=\frac{1}{2} \frac{n_{T}}{(1+\zeq)^{3}} \times \lim _{ dt\rightarrow 0 }
\nonumber \\
&& { \frac{G\(t+dt,m_i,m_j,m_l,k_i,k_j,k_l\)-G\(t,m_i,m_j,m_l,k_i,k_j,k_l\)}{dt}},
\nonumber \\
\n
where the factor $1/2$ accounts for that each merger event involves two PBHs.
The merger rate distribution of PBH binaries with charges are given by
\m
\label{R1}
\RR(t,m_i,m_j)=\int dm_ldk_i dk_j dk_l \RR(t,m_i,m_j,m_l,k_i,k_j,k_l) .
\notag \\
\n

In this section, we develop a formalism to calculate the merger rate distribution of PBH binaries with charges and a general mass function by taking into account gravitational torque and electromagnetic torque by the nearest primordial black hole. It is suitable for any PBH masses and charges. For a given probability distribution function of PBH masses and charges $P(m,k)$, we can  get the merger rate distribution of PBH with charges. In the next section, we will apply the formalism to the extremal charged case.

\section{Extremal charged PBH binaries}\label{Cases}
Ref \cite{Hawking:1974sw}  showed that a black hole evaporates or radiate as an ideal thermal blackbody and the temperature of black hole is only related to its surface gravity $\kappa$ via $T=\kappa/(2\pi)$, while $\kappa$ only depends  on three parameters: mass $M$, electric charge $Q$, and angular momentum $L$.  A Schwarzschild black hole ($Q=L=0$) with mass $M$ less than $M_*\sim 5\times 10^{14}$g  has temperature $T = 1/(8\pi M)$ and a lifetime less than the age of the Universe. For a rotating black hole, the angular momentum is emitted much faster than energy, so a rapidly rotating black hole will quickly become a nearly non-rotating state before most of its mass has been given up \cite{Page:1976ki}. In this section, we focus on  PBHs with masses $M$ much smaller than $M_*$. Because of Hawing radiation, those PBHs will quickly become extremal charged black holes. Their mass function could be arbitrary and only depend on their initial charge distribution. Thus we choose the probability distribution function of PBH masses and charges $P(m, k)$ as
\m
P(m,k)=\frac{\delta(k-1)+\delta(k+1)}{2} P(m).
\n
Only two PBHs with opposite charge ($\la=-1$) could form a bound system. The semi-major axis $a$, the dimensionless angular momentum $j$ and coalescence time $\tau$ of the formed binary can be approximated as
\m
a \approx 0.1 \frac{4 \pi \rhoeq x^4}{3 m_b},
\n
\m
j \approx \frac{x^2m_l \sin \theta}{y^2 m_b},
\n
\m
\tau \approx \frac{  a^3  j^{5}}{8 G^2 m_i m_j }.
\n
Applying the formalism in Sec.~\ref{Formalism}, the merger rate distribution of PBH binaries with charges are given by
\m
\RR(t,m_i,m_j)=\int dm_l \RR(t,m_i,m_j,m_l) .
\n
where
\m
&\RR&(t,m_i,m_j,m_l)
 \approx P\(m_i\) P\(m_j\) P\(m_l\)
\nonumber \\
& \times& 4.69\times 10^{6} \({ \Msun}\)^{{19 \over 22}}  \({ m_i m_j}\)^{-{19 \over 22}}
 \({ m_l}\)^{-{37 \over 22}} \({ m_{\rm pbh}}\)^{{16 \over 11}} \fpbh^{17 \over 11}
 \nonumber \\
& \times& \({t\over t_0}\)^{-{19\over 22}} \({ m_i+m_j}\)^{{12 \over 11}}.
\n
which can be interpreted as the merger rate density in unit of Gpc$^{-3}$ yr$^{-1}$$\Msun^{-2}$.
So,  $\alpha=-(m_i+m_j)^2\p^2 \ln \mathcal{R}(t,m_i,m_j)/\p m_i \p m_j=12/11$, which is independent of the PBH mass function. By contrast, for uncharged PBH binaries, $\alpha=36/37$ derived in \cite{Raidal:2018bbj,Liu:2018ess}.

In this section, we introduced a possible scenario to produce extremal charged PBHs with masses much smaller than $M_*\sim 5\times 10^{14}$g and worked out the merger rate distribution of those extremal charged PBHs. Those results can be used to test extremal charged PBH as DM. And those much lighter black holes can not be of stellar origin. Moreover, the result is also valid for PBHs with mass heavier than solar mass. For those extremal solar mass PBHs with dark electric charges which have a hidden U(1) symmetry and are formally described by the same Maxwell theory \cite{Cardoso_2016}, our result also can be used to distinguish uncharged PBHs and extremal charged PBHs.

\section{Conclusions and discussions}
\label{Conclusion}

   We have calculated gravitational radiation and electromagnetic radiation from point masses with charges in a Keplerian orbit and applied the result to work out the merger rate distribution of PBH binaries with charges and a general mass function by taking into account  gravitational torque and electromagnetic torque by the nearest PBH. For the extremal charged case, we find that $\alpha=-(m_i+m_j)^2\p^2 \ln {\cal R}(m_i,m_j)/\p m_i \p m_j=12/11$, which is independent of the mass function. PBHs are a natural DM candidate without requiring physics beyond the standard model. There are many constraints for uncharged PBH as all DM (see reviews \cite{Carr:2009jm, Carr:2020gox}). However, there are no observational constraints for such light and stable extremal charged PBHs with mass much smaller than $M_*\sim 5\times 10^{14}$ g as all DM. Such extremal charged black hole could be tested as DM. Two extremal charged PBHs could form a bound system only when they carry opposite charges. These binaries have $\alpha = 12/11$.
When they merger, they burst gamma rays due to annihilation of charges and become a non-extremal charged BH, which triggers striking Hawking radiation. The gamma rays during the merger and thereafter Hawking radiation from the non-extremal black hole may be detected by future observations.
   

In our calculation, we assumed that the spatial distribution of PBHs is random one. An additional consideration in calculating the merger rate is the cluster of PBHs which could considerably change the merger rate  \cite{Clesse:2016vqa,Desjacques:2018wuu,Bringmann:2018mxj,Suyama:2019cst}. This is an interesting topic, but it is believed that, for Gaussian initial conditions, the spatial distribution of PBHs is Poisson distributed with no additional clustering.

Finally, we discussed physical properties of charged black holes and their formation scenarios. In an asymptotically flat spacetime the Einstein-Maxwell theory has the Kerr-Newman black hole with an angular momentum $L=jM$ and an electric charge $Q$ and/or magnetic charge $P$ as well as a mass $M$. The extremal condition of $M^2 = j^2 + (Q^2+P^2)$ is achieved for a non-rotating black hole with electric charge when $M = Q$.
The hydrostatic equilibrium for multicomponent fluids with charges allows segregation of charges and gravitational collapse to form pairs of Kerr-Newman black holes with opposite charges \cite{Olson:1975ts,1978ApJ...220..743B,Ray:2003gt,Zhang:2016rli,Zajacek:2019kla} from high density plasma in the early universe. Then, a Kerr-Newman black hole with mass $M$ less than $M_*$ can quickly lose its angular momentum, become a Reissner-Nordstr\"{o}m black hole, and finally become an extremal charged black hole  through  Hawing radiation~\cite{Page:1976ki}. Gibbons argued \cite{Gibbons:1976sm} that black holes with electric and magnetic charges end up as extremal ones. However, it has recently been shown that even extremal charged black holes emit charged particle through Schwinger mechanism unless the BF bound holds which leads to Planck sized black holes stable against both the Schwinger effect and Hawking radiation~\cite{Chen:2017mnm}. Such an extremal charged black hole is stable and may be a candidate of DM. When they merge, the GWs signals are too weak to be detected by the present and near future observatories. However, the burst of gamma rays during the merger and the prompt Hawking radiation thereafter may be detected by telescopes. Our calculation may be used to test extremal charged PBHs as a dark mater candidate. The charged black hole has lots of implications to astrophysics and cosmology \cite{Kim:2019joy}. We have not included the emission of charges from extremal black holes, which goes beyond the scope of this paper and requires a further study.

In the early universe with a large Hubble constant, non-rotating charged black holes have both the event horizon and the cosmological horizon, which depend on the de Sitter radius, and the effect of the de Sitter space on Hawking radiation cannot be neglected~\cite{Romans:1991nq,Cai:1997ih}. The Schwinger mechanism from charged black holes in de Sitter space also differs from that of charged black holes in the asymptotically flat spacetime in that the Hubble radius affects the effective temperature for Schwinger mechanism and the emission of charges from the cosmological horizon affects that from the event horizon itself~\cite{Montero:2019ekk}. The detailed description requires a quantitative study.

Another scenario beyond the standard model for extremal PBHs is the dark quantum electrodynamics with dark electrons and photons, whose suppressed Schwinger effect gives the life time of PBHs longer than the age of the universe~\cite{Bai:2019zcd}. Still another scenario is the formation of black holes from gauge fields during the inflation~\cite{Maleknejad:2012fw}. We leave all these topics for future works.

\section*{Acknowledgement}
We would like to thank {\O}yvind Christiansen
for useful discussions. This work is supported in part by the National Natural Science Foundation of China Grants
 No.11690021, No.11690022, No.11851302, No.11821505, and No.11947302,
 in  part by the Strategic Priority Research Program of the Chinese Academy of Sciences Grant No. XDB23030100,
No. XDA15020701 and by Key Research Program of Frontier Sciences, CAS.

\appendix

\subsection*{Appendix}
\subsection{Angular Momentum Emission}
\label{AngularMomentum}
The rate of angular momentum emission due to electromagnetic waves is
\m
\frac{d J_{EM}^{i}}{d t}=- \int r^{2} d \Omega (-\epsilon^{i k l}\left(\partial_{0} A_{j}\right) x^{k} \partial^{l} A_{j}+\epsilon^{i k l} A_{k} \partial_{0} A_{l}).
\notag \\
\n
Here, we introduce a general direction dipole as given by
\m
p^{\prime i}(\mathbf{n})=P_{i j}(\mathbf{n}) p^{j}(t-r)
\n
where $P_{i j}=\delta_{i j}-n_{i} n_{j}$ is a projection operator that enforces the transversal gauge in the $\mathbf{n}$-direction. Thus, we can get
\m
A_i\cong \frac{P_{i j} p_{j}}{4\pi r}
\n
We have $\partial^{i} f(t-r)=-n^{i} \dot{f}$ and $\epsilon^{i k l} n^k n^l=0$, so for the orbital angular momentum term, we have
\m
\dot{L}_{EM}^{i}=\frac{\epsilon^{i k l}}{(4 \pi)^2} r \ddot{p}_{a} \dot{p}_{d} \int d \Omega P^{b a} n^{k} \partial^{l} P^{b d}.
\n
By using
\m
\partial^{l} P^{b d}=-\frac{1}{r} P^{l m}\left(\delta^{d m} n^{b}+\delta^{b m} n^{d}\right),
\n
\m
\int d \Omega n^{d} n^{k}=\frac{4}{3} \pi \delta^{d k},
\n
we can get
\m
\dot{L}_{EM}^{i}=-\frac{\epsilon^{i k l}}{12\pi}\dot{p}_k \ddot{p}_l.
\n
For the spin contribution, similarly, we have
\m
\dot{S}_{EM}^{i}=\frac{-\epsilon^{i k l}}{(4 \pi)^2} \ddot{p}_{a} \dot{p}_{d} \int d \Omega P^{ a l}  P^{d k}
=-\frac{\epsilon^{i k l}}{12\pi}\dot{p}_k \ddot{p}_l.
\n
Finally, we obtain
\m
\frac{d J_{EM}^{i}}{d t}= -\frac{\epsilon^{i k l}}{6\pi}\dot{p}_k \ddot{p}_l.
\n

\subsection{The coalescence time}
The total rate of energy and angular momentum emission due to electromagnetic radiation and gravitational radiation are given by
\m
\left\langle\frac{dE}{dt}\right\rangle=\left\langle\frac{dE_{EM}}{dt}\right\rangle+\left\langle\frac{dE_{GW}}{dt}\right\rangle,
\n
According to \eqref{E1} and \eqref{TL}, we have
\m
\label{dadt}
\frac{da}{dt}&=&-\frac{2 \left(37 e^4+292 e^2+96\right) G^3 (1-\la)^2 m_1 m_2 (m_1+m_2)}{15 a^3  \left(1-e^2\right)^{7/2}}
 \notag \\
&-& \frac{\left(e^2+2\right) G (1-\la) (m_2 Q_1-m_1 Q_2)^2}{6 \pi  a^2 \left(1-e^2\right)^{5/2} m_1 m_2},
\n

\m
\label{dedt}
\frac{de}{dt}&=& -\frac{e \left(121 e^2+304\right) G^3 (1-\la)^2 m_1 m_2 (m_1+m_2)}{15 a^4 c^5 \left(1-e^2\right)^{5/2}}
 \notag \\
&-& \frac{e G (1-\la) (m_2 Q_1-m_1 Q_2)^2}{4 \pi  a^3 \left(1-e^2\right)^{3/2} m_1 m_2}.
\n
If electromagnetic radiation is dominated, \eqref{dadt} and \eqref{dedt} become
\m
\label{dadt1}
\frac{da}{dt}=- \frac{\left(e^2+2\right) G (1-\la) (m_2 Q_1-m_1 Q_2)^2}{6 \pi  a^2 \left(1-e^2\right)^{5/2} m_1 m_2},
\n
\m
\label{dedt1}
\frac{de}{dt}=-\frac{e G (1-\la) (m_2 Q_1-m_1 Q_2)^2}{4 \pi  a^3 \left(1-e^2\right)^{3/2} m_1 m_2}.
\n
So, we can get
\m
\label{dade1}
\frac{da}{de} = \frac{2 a e^2+4 a}{3 e-3 e^3}.
\n
This equation can be integrated analytically, and gives
\m
a= \frac{c_0 e^{4/3}}{1-e^2},
\n
where $c_{0}$ is determined by the initial condition $a=a_{0}$ when $e=e_{0}$. We now compute the time to coalescence, $\tau_{EM}\left(a_{0}, e_{0}\right)$, of a binary system that, at an initial time $t_{0}=0,$ has semi-major axis $a_{0}$ and eccentricity $e_{0}$. When $e_{0}=0,$ we found
\m
\frac{da}{dt}=-\frac{ G (1-\la) (m_2 Q_1-m_1 Q_2)^2}{3 \pi  a^2  m_1 m_2},
\n
\m
\tau_{EM}\left(a_{0}, e_{0}=0\right)=\frac{\pi  a_0^3  m_1 m_2}{G (1-\lambda ) (m_2 Q_1-m_1 Q_2)^2}.
\n
For an elliptic orbit, we can integrate \eqref{dadt1} requiring $a(t)=0$ at $t=\tau_{EM}\left(a_{0}, e_{0}\right)$ or, equivalently, we can integrate \eqref{dedt1} requiring $e(t)=0$ at $t=\tau_{EM}\left(a_{0}, e_{0}\right)$, since we have seen that at the coalescence $e$ goes to zero. Since the analytic expression for $a(e)$ is simpler than the form of the inverse function $e(a)$, it is in fact better to use \eqref{dade1}, so we get
\m
\int_{0}^{\tau_{EM}\left(a_{0}, e_{0}\right)}dt&=&-\frac{4\pi m_1 m_2 }{G(1-\la)(m_2 Q_1-m_1 Q_2)^2}
\notag \\
&\times& \int_{e_{0}}^{0} d e \frac{a^{3}(e)\left(1-e^{2}\right)^{3 / 2}}{e},
\n
\m
\tau_{EM}\left(a_{0}, e_{0}\right)=\frac{4 \pi  \left(\frac{2-e_0^2}{\sqrt{1-e_0^2}}-2\right) m_1 m_2 a_0^3\left(1- e_0^2\right)^3}{G (1-\la) (m_2 Q_1-m_1 Q_2)^2}.
\notag \\
\n
For $e_0 \simeq 1$, we have
\m
\tau_{EM}\left(a_{0}, e_{0}\right) \approx \frac{4 \pi  a_0^3 \left(1-e_0^2\right)^{5/2} m_1 m_2}{G (1-\lambda ) (m_2 Q_1-m_1 Q_2)^2}.
\n
If gravitational radiation is dominated, \eqref{dadt} and \eqref{dedt} become
\m
\label{dadt2}
\frac{da}{dt}=-\frac{2 \left(37 e^4+292 e^2+96\right) G^3 (1-\la)^2 m_1 m_2 (m_1+m_2)}{15 a^3  \left(1-e^2\right)^{7/2}},
\notag \\
\n
\m
\label{dedt2}
\frac{de}{dt}=-\frac{e \left(121 e^2+304\right) G^3 (1-\la)^2 m_1 m_2 (m_1+m_2)}{15 a^4 c^5 \left(1-e^2\right)^{5/2}}.
\notag \\
\n
So, we can get
\m
\label{dade2}
\frac{da}{de} = -\frac{2 a \left(37 e^4+292 e^2+96\right)}{e \left(121 e^4+183 e^2-304\right)},
\n
\m
a(e)=\frac{c_0 e^{12/19} \left(121 e^2+304\right)^{870/2299}}{1-e^2}.
\n
When $e_{0}=0,$ we found
\m
\frac{da}{dt}=-\frac{64 G^3 (k+1)^2 m_1 m_2 (m_1+m_2)}{5 a^3},
\n
\m
\tau_{GW}\left(a_{0}, e_{0}=0\right)=\frac{5 a_0^4 }{256 G^3 (1-\la)^2 m_1 m_2 (m_1+m_2)}.
\notag \\
\n
For $e_0 \simeq 1$, we have
\m
\tau_{GW}\left(a_{0}, e_{0}\right) \approx \frac{3 a_0^4 \left(1-e_0^2\right)^{7/2} }{85 G^3 (1-\lambda )^2 m_1 m_2 (m_1+m_2)}.
\n


\bibliography{merger_v11}

\begin{thebibliography}{82}%
\makeatletter
\providecommand \@ifxundefined [1]{%
 \@ifx{#1\undefined}
}%
\providecommand \@ifnum [1]{%
 \ifnum #1\expandafter \@firstoftwo
 \else \expandafter \@secondoftwo
 \fi
}%
\providecommand \@ifx [1]{%
 \ifx #1\expandafter \@firstoftwo
 \else \expandafter \@secondoftwo
 \fi
}%
\providecommand \natexlab [1]{#1}%
\providecommand \enquote  [1]{``#1''}%
\providecommand \bibnamefont  [1]{#1}%
\providecommand \bibfnamefont [1]{#1}%
\providecommand \citenamefont [1]{#1}%
\providecommand \href@noop [0]{\@secondoftwo}%
\providecommand \href [0]{\begingroup \@sanitize@url \@href}%
\providecommand \@href[1]{\@@startlink{#1}\@@href}%
\providecommand \@@href[1]{\endgroup#1\@@endlink}%
\providecommand \@sanitize@url [0]{\catcode `\\12\catcode `\$12\catcode
  `\&12\catcode `\#12\catcode `\^12\catcode `\_12\catcode `\%12\relax}%
\providecommand \@@startlink[1]{}%
\providecommand \@@endlink[0]{}%
\providecommand \url  [0]{\begingroup\@sanitize@url \@url }%
\providecommand \@url [1]{\endgroup\@href {#1}{\urlprefix }}%
\providecommand \urlprefix  [0]{URL }%
\providecommand \Eprint [0]{\href }%
\providecommand \doibase [0]{http://dx.doi.org/}%
\providecommand \selectlanguage [0]{\@gobble}%
\providecommand \bibinfo  [0]{\@secondoftwo}%
\providecommand \bibfield  [0]{\@secondoftwo}%
\providecommand \translation [1]{[#1]}%
\providecommand \BibitemOpen [0]{}%
\providecommand \bibitemStop [0]{}%
\providecommand \bibitemNoStop [0]{.\EOS\space}%
\providecommand \EOS [0]{\spacefactor3000\relax}%
\providecommand \BibitemShut  [1]{\csname bibitem#1\endcsname}%
\let\auto@bib@innerbib\@empty
\bibitem [{\citenamefont {Hawking}(1971)}]{Hawking:1971ei}%
  \BibitemOpen
  \bibfield  {author} {\bibinfo {author} {\bibfnamefont {S.}~\bibnamefont
  {Hawking}},\ }\href@noop {} {\bibfield  {journal} {\bibinfo  {journal} {Mon.
  Not. Roy. Astron. Soc.}\ }\textbf {\bibinfo {volume} {152}},\ \bibinfo
  {pages} {75} (\bibinfo {year} {1971})}\BibitemShut {NoStop}%
\bibitem [{\citenamefont {Carr}\ and\ \citenamefont
  {Hawking}(1974)}]{Carr:1974nx}%
  \BibitemOpen
  \bibfield  {author} {\bibinfo {author} {\bibfnamefont {B.~J.}\ \bibnamefont
  {Carr}}\ and\ \bibinfo {author} {\bibfnamefont {S.~W.}\ \bibnamefont
  {Hawking}},\ }\href@noop {} {\bibfield  {journal} {\bibinfo  {journal} {Mon.
  Not. Roy. Astron. Soc.}\ }\textbf {\bibinfo {volume} {168}},\ \bibinfo
  {pages} {399} (\bibinfo {year} {1974})}\BibitemShut {NoStop}%
\bibitem [{\citenamefont {Carr}(1975)}]{Carr:1975qj}%
  \BibitemOpen
  \bibfield  {author} {\bibinfo {author} {\bibfnamefont {B.~J.}\ \bibnamefont
  {Carr}},\ }\href {\doibase 10.1086/153853} {\bibfield  {journal} {\bibinfo
  {journal} {Astrophys. J.}\ }\textbf {\bibinfo {volume} {201}},\ \bibinfo
  {pages} {1} (\bibinfo {year} {1975})}\BibitemShut {NoStop}%
\bibitem [{\citenamefont {Stewart}(1997{\natexlab{a}})}]{Stewart:1996ey}%
  \BibitemOpen
  \bibfield  {author} {\bibinfo {author} {\bibfnamefont {E.~D.}\ \bibnamefont
  {Stewart}},\ }\href {\doibase 10.1016/S0370-2693(96)01458-X} {\bibfield
  {journal} {\bibinfo  {journal} {Phys. Lett.}\ }\textbf {\bibinfo {volume}
  {B391}},\ \bibinfo {pages} {34} (\bibinfo {year} {1997}{\natexlab{a}})},\
  \Eprint {http://arxiv.org/abs/hep-ph/9606241} {arXiv:hep-ph/9606241 [hep-ph]}
  \BibitemShut {NoStop}%
\bibitem [{\citenamefont {Stewart}(1997{\natexlab{b}})}]{Stewart:1997wg}%
  \BibitemOpen
  \bibfield  {author} {\bibinfo {author} {\bibfnamefont {E.~D.}\ \bibnamefont
  {Stewart}},\ }\href {\doibase 10.1103/PhysRevD.56.2019} {\bibfield  {journal}
  {\bibinfo  {journal} {Phys. Rev.}\ }\textbf {\bibinfo {volume} {D56}},\
  \bibinfo {pages} {2019} (\bibinfo {year} {1997}{\natexlab{b}})},\ \Eprint
  {http://arxiv.org/abs/hep-ph/9703232} {arXiv:hep-ph/9703232 [hep-ph]}
  \BibitemShut {NoStop}%
\bibitem [{\citenamefont {Leach}\ \emph {et~al.}(2000)\citenamefont {Leach},
  \citenamefont {Grivell},\ and\ \citenamefont {Liddle}}]{Leach:2000ea}%
  \BibitemOpen
  \bibfield  {author} {\bibinfo {author} {\bibfnamefont {S.~M.}\ \bibnamefont
  {Leach}}, \bibinfo {author} {\bibfnamefont {I.~J.}\ \bibnamefont {Grivell}},
  \ and\ \bibinfo {author} {\bibfnamefont {A.~R.}\ \bibnamefont {Liddle}},\
  }\href {\doibase 10.1103/PhysRevD.62.043516} {\bibfield  {journal} {\bibinfo
  {journal} {Phys. Rev.}\ }\textbf {\bibinfo {volume} {D62}},\ \bibinfo {pages}
  {043516} (\bibinfo {year} {2000})},\ \Eprint
  {http://arxiv.org/abs/astro-ph/0004296} {arXiv:astro-ph/0004296 [astro-ph]}
  \BibitemShut {NoStop}%
\bibitem [{\citenamefont {Cheng}\ \emph {et~al.}(2018)\citenamefont {Cheng},
  \citenamefont {Lee},\ and\ \citenamefont {Ng}}]{Cheng:2018yyr}%
  \BibitemOpen
  \bibfield  {author} {\bibinfo {author} {\bibfnamefont {S.-L.}\ \bibnamefont
  {Cheng}}, \bibinfo {author} {\bibfnamefont {W.}~\bibnamefont {Lee}}, \ and\
  \bibinfo {author} {\bibfnamefont {K.-W.}\ \bibnamefont {Ng}},\ }\href
  {\doibase 10.1088/1475-7516/2018/07/001} {\bibfield  {journal} {\bibinfo
  {journal} {JCAP}\ }\textbf {\bibinfo {volume} {1807}},\ \bibinfo {pages}
  {001} (\bibinfo {year} {2018})},\ \Eprint {http://arxiv.org/abs/1801.09050}
  {arXiv:1801.09050 [astro-ph.CO]} \BibitemShut {NoStop}%
\bibitem [{\citenamefont {Gao}\ and\ \citenamefont {Guo}(2018)}]{Gao:2018pvq}%
  \BibitemOpen
  \bibfield  {author} {\bibinfo {author} {\bibfnamefont {T.-J.}\ \bibnamefont
  {Gao}}\ and\ \bibinfo {author} {\bibfnamefont {Z.-K.}\ \bibnamefont {Guo}},\
  }\href {\doibase 10.1103/PhysRevD.98.063526} {\bibfield  {journal} {\bibinfo
  {journal} {Phys. Rev.}\ }\textbf {\bibinfo {volume} {D98}},\ \bibinfo {pages}
  {063526} (\bibinfo {year} {2018})},\ \Eprint
  {http://arxiv.org/abs/1806.09320} {arXiv:1806.09320 [hep-ph]} \BibitemShut
  {NoStop}%
\bibitem [{\citenamefont {Deng}\ \emph {et~al.}(2017)\citenamefont {Deng},
  \citenamefont {Garriga},\ and\ \citenamefont {Vilenkin}}]{Deng:2016vzb}%
  \BibitemOpen
  \bibfield  {author} {\bibinfo {author} {\bibfnamefont {H.}~\bibnamefont
  {Deng}}, \bibinfo {author} {\bibfnamefont {J.}~\bibnamefont {Garriga}}, \
  and\ \bibinfo {author} {\bibfnamefont {A.}~\bibnamefont {Vilenkin}},\ }\href
  {\doibase 10.1088/1475-7516/2017/04/050} {\bibfield  {journal} {\bibinfo
  {journal} {JCAP}\ }\textbf {\bibinfo {volume} {1704}},\ \bibinfo {pages}
  {050} (\bibinfo {year} {2017})},\ \Eprint {http://arxiv.org/abs/1612.03753}
  {arXiv:1612.03753 [gr-qc]} \BibitemShut {NoStop}%
\bibitem [{\citenamefont {Liu}\ \emph {et~al.}(2019{\natexlab{a}})\citenamefont
  {Liu}, \citenamefont {Guo},\ and\ \citenamefont {Cai}}]{Liu:2019lul}%
  \BibitemOpen
  \bibfield  {author} {\bibinfo {author} {\bibfnamefont {J.}~\bibnamefont
  {Liu}}, \bibinfo {author} {\bibfnamefont {Z.-K.}\ \bibnamefont {Guo}}, \ and\
  \bibinfo {author} {\bibfnamefont {R.-G.}\ \bibnamefont {Cai}},\ }\href@noop
  {} {\  (\bibinfo {year} {2019}{\natexlab{a}})},\ \Eprint
  {http://arxiv.org/abs/1908.02662} {arXiv:1908.02662 [astro-ph.CO]}
  \BibitemShut {NoStop}%
\bibitem [{\citenamefont {Kodama}\ \emph {et~al.}(1982)\citenamefont {Kodama},
  \citenamefont {Sasaki},\ and\ \citenamefont {Sato}}]{Kodama:1982sf}%
  \BibitemOpen
  \bibfield  {author} {\bibinfo {author} {\bibfnamefont {H.}~\bibnamefont
  {Kodama}}, \bibinfo {author} {\bibfnamefont {M.}~\bibnamefont {Sasaki}}, \
  and\ \bibinfo {author} {\bibfnamefont {K.}~\bibnamefont {Sato}},\ }\href
  {\doibase 10.1143/PTP.68.1979} {\bibfield  {journal} {\bibinfo  {journal}
  {Prog. Theor. Phys.}\ }\textbf {\bibinfo {volume} {68}},\ \bibinfo {pages}
  {1979} (\bibinfo {year} {1982})}\BibitemShut {NoStop}%
\bibitem [{\citenamefont {Hawking}\ \emph {et~al.}(1982)\citenamefont
  {Hawking}, \citenamefont {Moss},\ and\ \citenamefont
  {Stewart}}]{Hawking:1982ga}%
  \BibitemOpen
  \bibfield  {author} {\bibinfo {author} {\bibfnamefont {S.~W.}\ \bibnamefont
  {Hawking}}, \bibinfo {author} {\bibfnamefont {I.~G.}\ \bibnamefont {Moss}}, \
  and\ \bibinfo {author} {\bibfnamefont {J.~M.}\ \bibnamefont {Stewart}},\
  }\href {\doibase 10.1103/PhysRevD.26.2681} {\bibfield  {journal} {\bibinfo
  {journal} {Phys. Rev.}\ }\textbf {\bibinfo {volume} {D26}},\ \bibinfo {pages}
  {2681} (\bibinfo {year} {1982})}\BibitemShut {NoStop}%
\bibitem [{\citenamefont {Lewicki}\ and\ \citenamefont
  {Vaskonen}(2019)}]{Lewicki:2019gmv}%
  \BibitemOpen
  \bibfield  {author} {\bibinfo {author} {\bibfnamefont {M.}~\bibnamefont
  {Lewicki}}\ and\ \bibinfo {author} {\bibfnamefont {V.}~\bibnamefont
  {Vaskonen}},\ }\href@noop {} {\  (\bibinfo {year} {2019})},\ \Eprint
  {http://arxiv.org/abs/1912.00997} {arXiv:1912.00997 [astro-ph.CO]}
  \BibitemShut {NoStop}%
\bibitem [{\citenamefont {Martin}\ \emph {et~al.}(2019)\citenamefont {Martin},
  \citenamefont {Papanikolaou},\ and\ \citenamefont {Vennin}}]{Martin:2019nuw}%
  \BibitemOpen
  \bibfield  {author} {\bibinfo {author} {\bibfnamefont {J.}~\bibnamefont
  {Martin}}, \bibinfo {author} {\bibfnamefont {T.}~\bibnamefont
  {Papanikolaou}}, \ and\ \bibinfo {author} {\bibfnamefont {V.}~\bibnamefont
  {Vennin}},\ }\href@noop {} {\  (\bibinfo {year} {2019})},\ \Eprint
  {http://arxiv.org/abs/1907.04236} {arXiv:1907.04236 [astro-ph.CO]}
  \BibitemShut {NoStop}%
\bibitem [{\citenamefont {Cai}\ \emph {et~al.}(2018{\natexlab{a}})\citenamefont
  {Cai}, \citenamefont {Tong}, \citenamefont {Wang},\ and\ \citenamefont
  {Yan}}]{Cai:2018tuh}%
  \BibitemOpen
  \bibfield  {author} {\bibinfo {author} {\bibfnamefont {Y.-F.}\ \bibnamefont
  {Cai}}, \bibinfo {author} {\bibfnamefont {X.}~\bibnamefont {Tong}}, \bibinfo
  {author} {\bibfnamefont {D.-G.}\ \bibnamefont {Wang}}, \ and\ \bibinfo
  {author} {\bibfnamefont {S.-F.}\ \bibnamefont {Yan}},\ }\href {\doibase
  10.1103/PhysRevLett.121.081306} {\bibfield  {journal} {\bibinfo  {journal}
  {Phys. Rev. Lett.}\ }\textbf {\bibinfo {volume} {121}},\ \bibinfo {pages}
  {081306} (\bibinfo {year} {2018}{\natexlab{a}})},\ \Eprint
  {http://arxiv.org/abs/1805.03639} {arXiv:1805.03639 [astro-ph.CO]}
  \BibitemShut {NoStop}%
\bibitem [{\citenamefont {Cai}\ \emph {et~al.}(2019{\natexlab{a}})\citenamefont
  {Cai}, \citenamefont {Guo}, \citenamefont {Liu}, \citenamefont {Liu},\ and\
  \citenamefont {Yang}}]{Cai:2019bmk}%
  \BibitemOpen
  \bibfield  {author} {\bibinfo {author} {\bibfnamefont {R.-G.}\ \bibnamefont
  {Cai}}, \bibinfo {author} {\bibfnamefont {Z.-K.}\ \bibnamefont {Guo}},
  \bibinfo {author} {\bibfnamefont {J.}~\bibnamefont {Liu}}, \bibinfo {author}
  {\bibfnamefont {L.}~\bibnamefont {Liu}}, \ and\ \bibinfo {author}
  {\bibfnamefont {X.-Y.}\ \bibnamefont {Yang}},\ }\href@noop {} {\  (\bibinfo
  {year} {2019}{\natexlab{a}})},\ \Eprint {http://arxiv.org/abs/1912.10437}
  {arXiv:1912.10437 [astro-ph.CO]} \BibitemShut {NoStop}%
\bibitem [{\citenamefont {Carr}\ \emph {et~al.}(2010)\citenamefont {Carr},
  \citenamefont {Kohri}, \citenamefont {Sendouda},\ and\ \citenamefont
  {Yokoyama}}]{Carr:2009jm}%
  \BibitemOpen
  \bibfield  {author} {\bibinfo {author} {\bibfnamefont {B.~J.}\ \bibnamefont
  {Carr}}, \bibinfo {author} {\bibfnamefont {K.}~\bibnamefont {Kohri}},
  \bibinfo {author} {\bibfnamefont {Y.}~\bibnamefont {Sendouda}}, \ and\
  \bibinfo {author} {\bibfnamefont {J.}~\bibnamefont {Yokoyama}},\ }\href
  {\doibase 10.1103/PhysRevD.81.104019} {\bibfield  {journal} {\bibinfo
  {journal} {Phys. Rev.}\ }\textbf {\bibinfo {volume} {D81}},\ \bibinfo {pages}
  {104019} (\bibinfo {year} {2010})},\ \Eprint {http://arxiv.org/abs/0912.5297}
  {arXiv:0912.5297 [astro-ph.CO]} \BibitemShut {NoStop}%
\bibitem [{\citenamefont {Cai}\ \emph {et~al.}(2018{\natexlab{b}})\citenamefont
  {Cai}, \citenamefont {Liu},\ and\ \citenamefont {Wang}}]{Cai:2018rqf}%
  \BibitemOpen
  \bibfield  {author} {\bibinfo {author} {\bibfnamefont {R.-G.}\ \bibnamefont
  {Cai}}, \bibinfo {author} {\bibfnamefont {T.-B.}\ \bibnamefont {Liu}}, \ and\
  \bibinfo {author} {\bibfnamefont {S.-J.}\ \bibnamefont {Wang}},\ }\href
  {\doibase 10.1103/PhysRevD.98.043538} {\bibfield  {journal} {\bibinfo
  {journal} {Phys. Rev.}\ }\textbf {\bibinfo {volume} {D98}},\ \bibinfo {pages}
  {043538} (\bibinfo {year} {2018}{\natexlab{b}})},\ \Eprint
  {http://arxiv.org/abs/1806.05390} {arXiv:1806.05390 [astro-ph.CO]}
  \BibitemShut {NoStop}%
\bibitem [{\citenamefont {Belotsky}\ \emph {et~al.}(2018)\citenamefont
  {Belotsky}, \citenamefont {Dokuchaev}, \citenamefont {Eroshenko},
  \citenamefont {Esipova}, \citenamefont {Khlopov}, \citenamefont {Khromykh},
  \citenamefont {Kirillov}, \citenamefont {Nikulin}, \citenamefont {Rubin},\
  and\ \citenamefont {Svadkovsky}}]{Belotsky:2018wph}%
  \BibitemOpen
  \bibfield  {author} {\bibinfo {author} {\bibfnamefont {K.~M.}\ \bibnamefont
  {Belotsky}}, \bibinfo {author} {\bibfnamefont {V.~I.}\ \bibnamefont
  {Dokuchaev}}, \bibinfo {author} {\bibfnamefont {Y.~N.}\ \bibnamefont
  {Eroshenko}}, \bibinfo {author} {\bibfnamefont {E.~A.}\ \bibnamefont
  {Esipova}}, \bibinfo {author} {\bibfnamefont {M.~{\relax Yu}.}\ \bibnamefont
  {Khlopov}}, \bibinfo {author} {\bibfnamefont {L.~A.}\ \bibnamefont
  {Khromykh}}, \bibinfo {author} {\bibfnamefont {A.~A.}\ \bibnamefont
  {Kirillov}}, \bibinfo {author} {\bibfnamefont {V.~V.}\ \bibnamefont
  {Nikulin}}, \bibinfo {author} {\bibfnamefont {S.~G.}\ \bibnamefont {Rubin}},
  \ and\ \bibinfo {author} {\bibfnamefont {I.~V.}\ \bibnamefont {Svadkovsky}},\
  }\href@noop {} {\  (\bibinfo {year} {2018})},\ \Eprint
  {http://arxiv.org/abs/1807.06590} {arXiv:1807.06590 [astro-ph.CO]}
  \BibitemShut {NoStop}%
\bibitem [{\citenamefont {Carr}\ \emph {et~al.}(2016)\citenamefont {Carr},
  \citenamefont {Kuhnel},\ and\ \citenamefont {Sandstad}}]{Carr:2016drx}%
  \BibitemOpen
  \bibfield  {author} {\bibinfo {author} {\bibfnamefont {B.}~\bibnamefont
  {Carr}}, \bibinfo {author} {\bibfnamefont {F.}~\bibnamefont {Kuhnel}}, \ and\
  \bibinfo {author} {\bibfnamefont {M.}~\bibnamefont {Sandstad}},\ }\href
  {\doibase 10.1103/PhysRevD.94.083504} {\bibfield  {journal} {\bibinfo
  {journal} {Phys. Rev.}\ }\textbf {\bibinfo {volume} {D94}},\ \bibinfo {pages}
  {083504} (\bibinfo {year} {2016})},\ \Eprint
  {http://arxiv.org/abs/1607.06077} {arXiv:1607.06077 [astro-ph.CO]}
  \BibitemShut {NoStop}%
\bibitem [{\citenamefont {Sasaki}\ \emph {et~al.}(2018)\citenamefont {Sasaki},
  \citenamefont {Suyama}, \citenamefont {Tanaka},\ and\ \citenamefont
  {Yokoyama}}]{Sasaki:2018dmp}%
  \BibitemOpen
  \bibfield  {author} {\bibinfo {author} {\bibfnamefont {M.}~\bibnamefont
  {Sasaki}}, \bibinfo {author} {\bibfnamefont {T.}~\bibnamefont {Suyama}},
  \bibinfo {author} {\bibfnamefont {T.}~\bibnamefont {Tanaka}}, \ and\ \bibinfo
  {author} {\bibfnamefont {S.}~\bibnamefont {Yokoyama}},\ }\href {\doibase
  10.1088/1361-6382/aaa7b4} {\bibfield  {journal} {\bibinfo  {journal} {Class.
  Quant. Grav.}\ }\textbf {\bibinfo {volume} {35}},\ \bibinfo {pages} {063001}
  (\bibinfo {year} {2018})},\ \Eprint {http://arxiv.org/abs/1801.05235}
  {arXiv:1801.05235 [astro-ph.CO]} \BibitemShut {NoStop}%
\bibitem [{\citenamefont {Cai}\ \emph {et~al.}(2019{\natexlab{b}})\citenamefont
  {Cai}, \citenamefont {Pi},\ and\ \citenamefont {Sasaki}}]{Cai:2018dig}%
  \BibitemOpen
  \bibfield  {author} {\bibinfo {author} {\bibfnamefont {R.-G.}\ \bibnamefont
  {Cai}}, \bibinfo {author} {\bibfnamefont {S.}~\bibnamefont {Pi}}, \ and\
  \bibinfo {author} {\bibfnamefont {M.}~\bibnamefont {Sasaki}},\ }\href
  {\doibase 10.1103/PhysRevLett.122.201101} {\bibfield  {journal} {\bibinfo
  {journal} {Phys. Rev. Lett.}\ }\textbf {\bibinfo {volume} {122}},\ \bibinfo
  {pages} {201101} (\bibinfo {year} {2019}{\natexlab{b}})},\ \Eprint
  {http://arxiv.org/abs/1810.11000} {arXiv:1810.11000 [astro-ph.CO]}
  \BibitemShut {NoStop}%
\bibitem [{\citenamefont {Kannike}\ \emph {et~al.}(2017)\citenamefont
  {Kannike}, \citenamefont {Marzola}, \citenamefont {Raidal},\ and\
  \citenamefont {Veermäe}}]{Kannike:2017bxn}%
  \BibitemOpen
  \bibfield  {author} {\bibinfo {author} {\bibfnamefont {K.}~\bibnamefont
  {Kannike}}, \bibinfo {author} {\bibfnamefont {L.}~\bibnamefont {Marzola}},
  \bibinfo {author} {\bibfnamefont {M.}~\bibnamefont {Raidal}}, \ and\ \bibinfo
  {author} {\bibfnamefont {H.}~\bibnamefont {Veermäe}},\ }\href {\doibase
  10.1088/1475-7516/2017/09/020} {\bibfield  {journal} {\bibinfo  {journal}
  {JCAP}\ }\textbf {\bibinfo {volume} {1709}},\ \bibinfo {pages} {020}
  (\bibinfo {year} {2017})},\ \Eprint {http://arxiv.org/abs/1705.06225}
  {arXiv:1705.06225 [astro-ph.CO]} \BibitemShut {NoStop}%
\bibitem [{\citenamefont {Kuhnel}\ and\ \citenamefont
  {Freese}(2019)}]{Kuhnel:2019xes}%
  \BibitemOpen
  \bibfield  {author} {\bibinfo {author} {\bibfnamefont {F.}~\bibnamefont
  {Kuhnel}}\ and\ \bibinfo {author} {\bibfnamefont {K.}~\bibnamefont
  {Freese}},\ }\href@noop {} {\  (\bibinfo {year} {2019})},\ \Eprint
  {http://arxiv.org/abs/1906.02744} {arXiv:1906.02744 [gr-qc]} \BibitemShut
  {NoStop}%
\bibitem [{\citenamefont {Gow}\ \emph {et~al.}(2019)\citenamefont {Gow},
  \citenamefont {Byrnes}, \citenamefont {Hall},\ and\ \citenamefont
  {Peacock}}]{Gow:2019pok}%
  \BibitemOpen
  \bibfield  {author} {\bibinfo {author} {\bibfnamefont {A.~D.}\ \bibnamefont
  {Gow}}, \bibinfo {author} {\bibfnamefont {C.~T.}\ \bibnamefont {Byrnes}},
  \bibinfo {author} {\bibfnamefont {A.}~\bibnamefont {Hall}}, \ and\ \bibinfo
  {author} {\bibfnamefont {J.~A.}\ \bibnamefont {Peacock}},\ }\href@noop {} {\
  (\bibinfo {year} {2019})},\ \Eprint {http://arxiv.org/abs/1911.12685}
  {arXiv:1911.12685 [astro-ph.CO]} \BibitemShut {NoStop}%
\bibitem [{\citenamefont {Chen}\ \emph
  {et~al.}(2018{\natexlab{a}})\citenamefont {Chen}, \citenamefont {Huang},\
  and\ \citenamefont {Huang}}]{Chen:2018rzo}%
  \BibitemOpen
  \bibfield  {author} {\bibinfo {author} {\bibfnamefont {Z.-C.}\ \bibnamefont
  {Chen}}, \bibinfo {author} {\bibfnamefont {F.}~\bibnamefont {Huang}}, \ and\
  \bibinfo {author} {\bibfnamefont {Q.-G.}\ \bibnamefont {Huang}},\ }\href@noop
  {} {\  (\bibinfo {year} {2018}{\natexlab{a}})},\ \Eprint
  {http://arxiv.org/abs/1809.10360} {arXiv:1809.10360 [gr-qc]} \BibitemShut
  {NoStop}%
\bibitem [{\citenamefont {Saito}\ and\ \citenamefont
  {Yokoyama}(2009)}]{Saito:2008jc}%
  \BibitemOpen
  \bibfield  {author} {\bibinfo {author} {\bibfnamefont {R.}~\bibnamefont
  {Saito}}\ and\ \bibinfo {author} {\bibfnamefont {J.}~\bibnamefont
  {Yokoyama}},\ }\href {\doibase 10.1103/PhysRevLett.102.161101,
  10.1103/PhysRevLett.107.069901} {\bibfield  {journal} {\bibinfo  {journal}
  {Phys. Rev. Lett.}\ }\textbf {\bibinfo {volume} {102}},\ \bibinfo {pages}
  {161101} (\bibinfo {year} {2009})},\ \bibinfo {note} {[Erratum: Phys. Rev.
  Lett.107,069901(2011)]},\ \Eprint {http://arxiv.org/abs/0812.4339}
  {arXiv:0812.4339 [astro-ph]} \BibitemShut {NoStop}%
\bibitem [{\citenamefont {Wang}\ \emph {et~al.}(2019)\citenamefont {Wang},
  \citenamefont {Terada},\ and\ \citenamefont {Kohri}}]{Wang:2019kaf}%
  \BibitemOpen
  \bibfield  {author} {\bibinfo {author} {\bibfnamefont {S.}~\bibnamefont
  {Wang}}, \bibinfo {author} {\bibfnamefont {T.}~\bibnamefont {Terada}}, \ and\
  \bibinfo {author} {\bibfnamefont {K.}~\bibnamefont {Kohri}},\ }\href
  {\doibase 10.1103/PhysRevD.99.103531} {\bibfield  {journal} {\bibinfo
  {journal} {Phys. Rev.}\ }\textbf {\bibinfo {volume} {D99}},\ \bibinfo {pages}
  {103531} (\bibinfo {year} {2019})},\ \Eprint
  {http://arxiv.org/abs/1903.05924} {arXiv:1903.05924 [astro-ph.CO]}
  \BibitemShut {NoStop}%
\bibitem [{\citenamefont {Chen}\ and\ \citenamefont
  {Huang}(2019)}]{Chen:2019irf}%
  \BibitemOpen
  \bibfield  {author} {\bibinfo {author} {\bibfnamefont {Z.-C.}\ \bibnamefont
  {Chen}}\ and\ \bibinfo {author} {\bibfnamefont {Q.-G.}\ \bibnamefont
  {Huang}},\ }\href@noop {} {\  (\bibinfo {year} {2019})},\ \Eprint
  {http://arxiv.org/abs/1904.02396} {arXiv:1904.02396 [astro-ph.CO]}
  \BibitemShut {NoStop}%
\bibitem [{\citenamefont {Laha}(2019)}]{Laha:2019ssq}%
  \BibitemOpen
  \bibfield  {author} {\bibinfo {author} {\bibfnamefont {R.}~\bibnamefont
  {Laha}},\ }\href {\doibase 10.1103/PhysRevLett.123.251101} {\bibfield
  {journal} {\bibinfo  {journal} {Phys. Rev. Lett.}\ }\textbf {\bibinfo
  {volume} {123}},\ \bibinfo {pages} {251101} (\bibinfo {year} {2019})},\
  \Eprint {http://arxiv.org/abs/1906.09994} {arXiv:1906.09994 [astro-ph.HE]}
  \BibitemShut {NoStop}%
\bibitem [{\citenamefont {Cai}\ \emph {et~al.}(2019{\natexlab{c}})\citenamefont
  {Cai}, \citenamefont {Pi}, \citenamefont {Wang},\ and\ \citenamefont
  {Yang}}]{Cai:2019elf}%
  \BibitemOpen
  \bibfield  {author} {\bibinfo {author} {\bibfnamefont {R.-G.}\ \bibnamefont
  {Cai}}, \bibinfo {author} {\bibfnamefont {S.}~\bibnamefont {Pi}}, \bibinfo
  {author} {\bibfnamefont {S.-J.}\ \bibnamefont {Wang}}, \ and\ \bibinfo
  {author} {\bibfnamefont {X.-Y.}\ \bibnamefont {Yang}},\ }\href {\doibase
  10.1088/1475-7516/2019/10/059} {\bibfield  {journal} {\bibinfo  {journal}
  {JCAP}\ }\textbf {\bibinfo {volume} {1910}},\ \bibinfo {pages} {059}
  (\bibinfo {year} {2019}{\natexlab{c}})},\ \bibinfo {note}
  {[JCAP1910,no.10,059(2019)]},\ \Eprint {http://arxiv.org/abs/1907.06372}
  {arXiv:1907.06372 [astro-ph.CO]} \BibitemShut {NoStop}%
\bibitem [{\citenamefont {Chen}\ \emph {et~al.}(2019)\citenamefont {Chen},
  \citenamefont {Yuan},\ and\ \citenamefont {Huang}}]{Chen:2019xse}%
  \BibitemOpen
  \bibfield  {author} {\bibinfo {author} {\bibfnamefont {Z.-C.}\ \bibnamefont
  {Chen}}, \bibinfo {author} {\bibfnamefont {C.}~\bibnamefont {Yuan}}, \ and\
  \bibinfo {author} {\bibfnamefont {Q.-G.}\ \bibnamefont {Huang}},\ }\href@noop
  {} {\  (\bibinfo {year} {2019})},\ \Eprint {http://arxiv.org/abs/1910.12239}
  {arXiv:1910.12239 [astro-ph.CO]} \BibitemShut {NoStop}%
\bibitem [{\citenamefont {Dasgupta}\ \emph {et~al.}(2019)\citenamefont
  {Dasgupta}, \citenamefont {Laha},\ and\ \citenamefont
  {Ray}}]{Dasgupta:2019cae}%
  \BibitemOpen
  \bibfield  {author} {\bibinfo {author} {\bibfnamefont {B.}~\bibnamefont
  {Dasgupta}}, \bibinfo {author} {\bibfnamefont {R.}~\bibnamefont {Laha}}, \
  and\ \bibinfo {author} {\bibfnamefont {A.}~\bibnamefont {Ray}},\ }\href@noop
  {} {\  (\bibinfo {year} {2019})},\ \Eprint {http://arxiv.org/abs/1912.01014}
  {arXiv:1912.01014 [hep-ph]} \BibitemShut {NoStop}%
\bibitem [{\citenamefont {Abbott}\ \emph {et~al.}(2016)\citenamefont {Abbott}
  \emph {et~al.}}]{Abbott:2016blz}%
  \BibitemOpen
  \bibfield  {author} {\bibinfo {author} {\bibfnamefont {B.~P.}\ \bibnamefont
  {Abbott}} \emph {et~al.} (\bibinfo {collaboration} {LIGO Scientific,
  Virgo}),\ }\href {\doibase 10.1103/PhysRevLett.116.061102} {\bibfield
  {journal} {\bibinfo  {journal} {Phys. Rev. Lett.}\ }\textbf {\bibinfo
  {volume} {116}},\ \bibinfo {pages} {061102} (\bibinfo {year} {2016})},\
  \Eprint {http://arxiv.org/abs/1602.03837} {arXiv:1602.03837 [gr-qc]}
  \BibitemShut {NoStop}%
\bibitem [{\citenamefont {Belczynski}\ \emph {et~al.}(2016)\citenamefont
  {Belczynski}, \citenamefont {Holz}, \citenamefont {Bulik},\ and\
  \citenamefont {O'Shaughnessy}}]{Belczynski:2016obo}%
  \BibitemOpen
  \bibfield  {author} {\bibinfo {author} {\bibfnamefont {K.}~\bibnamefont
  {Belczynski}}, \bibinfo {author} {\bibfnamefont {D.~E.}\ \bibnamefont
  {Holz}}, \bibinfo {author} {\bibfnamefont {T.}~\bibnamefont {Bulik}}, \ and\
  \bibinfo {author} {\bibfnamefont {R.}~\bibnamefont {O'Shaughnessy}},\ }\href
  {\doibase 10.1038/nature18322} {\bibfield  {journal} {\bibinfo  {journal}
  {Nature}\ }\textbf {\bibinfo {volume} {534}},\ \bibinfo {pages} {512}
  (\bibinfo {year} {2016})},\ \Eprint {http://arxiv.org/abs/1602.04531}
  {arXiv:1602.04531 [astro-ph.HE]} \BibitemShut {NoStop}%
\bibitem [{\citenamefont {Belczynski}\ \emph {et~al.}(2017)\citenamefont
  {Belczynski} \emph {et~al.}}]{Belczynski:2017gds}%
  \BibitemOpen
  \bibfield  {author} {\bibinfo {author} {\bibfnamefont {K.}~\bibnamefont
  {Belczynski}} \emph {et~al.},\ }\href@noop {} {\  (\bibinfo {year} {2017})},\
  \Eprint {http://arxiv.org/abs/1706.07053} {arXiv:1706.07053 [astro-ph.HE]}
  \BibitemShut {NoStop}%
\bibitem [{\citenamefont {Bird}\ \emph {et~al.}(2016)\citenamefont {Bird},
  \citenamefont {Cholis}, \citenamefont {Muñoz}, \citenamefont {Ali-Haïmoud},
  \citenamefont {Kamionkowski}, \citenamefont {Kovetz}, \citenamefont
  {Raccanelli},\ and\ \citenamefont {Riess}}]{Bird:2016dcv}%
  \BibitemOpen
  \bibfield  {author} {\bibinfo {author} {\bibfnamefont {S.}~\bibnamefont
  {Bird}}, \bibinfo {author} {\bibfnamefont {I.}~\bibnamefont {Cholis}},
  \bibinfo {author} {\bibfnamefont {J.~B.}\ \bibnamefont {Muñoz}}, \bibinfo
  {author} {\bibfnamefont {Y.}~\bibnamefont {Ali-Haïmoud}}, \bibinfo {author}
  {\bibfnamefont {M.}~\bibnamefont {Kamionkowski}}, \bibinfo {author}
  {\bibfnamefont {E.~D.}\ \bibnamefont {Kovetz}}, \bibinfo {author}
  {\bibfnamefont {A.}~\bibnamefont {Raccanelli}}, \ and\ \bibinfo {author}
  {\bibfnamefont {A.~G.}\ \bibnamefont {Riess}},\ }\href {\doibase
  10.1103/PhysRevLett.116.201301} {\bibfield  {journal} {\bibinfo  {journal}
  {Phys. Rev. Lett.}\ }\textbf {\bibinfo {volume} {116}},\ \bibinfo {pages}
  {201301} (\bibinfo {year} {2016})},\ \Eprint
  {http://arxiv.org/abs/1603.00464} {arXiv:1603.00464 [astro-ph.CO]}
  \BibitemShut {NoStop}%
\bibitem [{\citenamefont {Clesse}\ and\ \citenamefont
  {García-Bellido}(2017)}]{Clesse:2016vqa}%
  \BibitemOpen
  \bibfield  {author} {\bibinfo {author} {\bibfnamefont {S.}~\bibnamefont
  {Clesse}}\ and\ \bibinfo {author} {\bibfnamefont {J.}~\bibnamefont
  {García-Bellido}},\ }\href {\doibase 10.1016/j.dark.2016.10.002} {\bibfield
  {journal} {\bibinfo  {journal} {Phys. Dark Univ.}\ }\textbf {\bibinfo
  {volume} {15}},\ \bibinfo {pages} {142} (\bibinfo {year} {2017})},\ \Eprint
  {http://arxiv.org/abs/1603.05234} {arXiv:1603.05234 [astro-ph.CO]}
  \BibitemShut {NoStop}%
\bibitem [{\citenamefont {Nishikawa}\ \emph {et~al.}(2019)\citenamefont
  {Nishikawa}, \citenamefont {Kovetz}, \citenamefont {Kamionkowski},\ and\
  \citenamefont {Silk}}]{Nishikawa:2017chy}%
  \BibitemOpen
  \bibfield  {author} {\bibinfo {author} {\bibfnamefont {H.}~\bibnamefont
  {Nishikawa}}, \bibinfo {author} {\bibfnamefont {E.~D.}\ \bibnamefont
  {Kovetz}}, \bibinfo {author} {\bibfnamefont {M.}~\bibnamefont
  {Kamionkowski}}, \ and\ \bibinfo {author} {\bibfnamefont {J.}~\bibnamefont
  {Silk}},\ }\href {\doibase 10.1103/PhysRevD.99.043533} {\bibfield  {journal}
  {\bibinfo  {journal} {Phys. Rev.}\ }\textbf {\bibinfo {volume} {D99}},\
  \bibinfo {pages} {043533} (\bibinfo {year} {2019})},\ \Eprint
  {http://arxiv.org/abs/1708.08449} {arXiv:1708.08449 [astro-ph.CO]}
  \BibitemShut {NoStop}%
\bibitem [{\citenamefont {Nakamura}\ \emph {et~al.}(1997)\citenamefont
  {Nakamura}, \citenamefont {Sasaki}, \citenamefont {Tanaka},\ and\
  \citenamefont {Thorne}}]{Nakamura:1997sm}%
  \BibitemOpen
  \bibfield  {author} {\bibinfo {author} {\bibfnamefont {T.}~\bibnamefont
  {Nakamura}}, \bibinfo {author} {\bibfnamefont {M.}~\bibnamefont {Sasaki}},
  \bibinfo {author} {\bibfnamefont {T.}~\bibnamefont {Tanaka}}, \ and\ \bibinfo
  {author} {\bibfnamefont {K.~S.}\ \bibnamefont {Thorne}},\ }\href {\doibase
  10.1086/310886} {\bibfield  {journal} {\bibinfo  {journal} {Astrophys. J.}\
  }\textbf {\bibinfo {volume} {487}},\ \bibinfo {pages} {L139} (\bibinfo {year}
  {1997})},\ \Eprint {http://arxiv.org/abs/astro-ph/9708060}
  {arXiv:astro-ph/9708060 [astro-ph]} \BibitemShut {NoStop}%
\bibitem [{\citenamefont {Ioka}\ \emph {et~al.}(1998)\citenamefont {Ioka},
  \citenamefont {Chiba}, \citenamefont {Tanaka},\ and\ \citenamefont
  {Nakamura}}]{Ioka:1998nz}%
  \BibitemOpen
  \bibfield  {author} {\bibinfo {author} {\bibfnamefont {K.}~\bibnamefont
  {Ioka}}, \bibinfo {author} {\bibfnamefont {T.}~\bibnamefont {Chiba}},
  \bibinfo {author} {\bibfnamefont {T.}~\bibnamefont {Tanaka}}, \ and\ \bibinfo
  {author} {\bibfnamefont {T.}~\bibnamefont {Nakamura}},\ }\href {\doibase
  10.1103/PhysRevD.58.063003} {\bibfield  {journal} {\bibinfo  {journal} {Phys.
  Rev.}\ }\textbf {\bibinfo {volume} {D58}},\ \bibinfo {pages} {063003}
  (\bibinfo {year} {1998})},\ \Eprint {http://arxiv.org/abs/astro-ph/9807018}
  {arXiv:astro-ph/9807018 [astro-ph]} \BibitemShut {NoStop}%
\bibitem [{\citenamefont {Sasaki}\ \emph {et~al.}(2016)\citenamefont {Sasaki},
  \citenamefont {Suyama}, \citenamefont {Tanaka},\ and\ \citenamefont
  {Yokoyama}}]{Sasaki:2016jop}%
  \BibitemOpen
  \bibfield  {author} {\bibinfo {author} {\bibfnamefont {M.}~\bibnamefont
  {Sasaki}}, \bibinfo {author} {\bibfnamefont {T.}~\bibnamefont {Suyama}},
  \bibinfo {author} {\bibfnamefont {T.}~\bibnamefont {Tanaka}}, \ and\ \bibinfo
  {author} {\bibfnamefont {S.}~\bibnamefont {Yokoyama}},\ }\href {\doibase
  10.1103/PhysRevLett.121.059901, 10.1103/PhysRevLett.117.061101} {\bibfield
  {journal} {\bibinfo  {journal} {Phys. Rev. Lett.}\ }\textbf {\bibinfo
  {volume} {117}},\ \bibinfo {pages} {061101} (\bibinfo {year} {2016})},\
  \bibinfo {note} {[erratum: Phys. Rev. Lett.121,no.5,059901(2018)]},\ \Eprint
  {http://arxiv.org/abs/1603.08338} {arXiv:1603.08338 [astro-ph.CO]}
  \BibitemShut {NoStop}%
\bibitem [{\citenamefont {Raidal}\ \emph {et~al.}(2017)\citenamefont {Raidal},
  \citenamefont {Vaskonen},\ and\ \citenamefont {Veermäe}}]{Raidal:2017mfl}%
  \BibitemOpen
  \bibfield  {author} {\bibinfo {author} {\bibfnamefont {M.}~\bibnamefont
  {Raidal}}, \bibinfo {author} {\bibfnamefont {V.}~\bibnamefont {Vaskonen}}, \
  and\ \bibinfo {author} {\bibfnamefont {H.}~\bibnamefont {Veermäe}},\ }\href
  {\doibase 10.1088/1475-7516/2017/09/037} {\bibfield  {journal} {\bibinfo
  {journal} {JCAP}\ }\textbf {\bibinfo {volume} {1709}},\ \bibinfo {pages}
  {037} (\bibinfo {year} {2017})},\ \Eprint {http://arxiv.org/abs/1707.01480}
  {arXiv:1707.01480 [astro-ph.CO]} \BibitemShut {NoStop}%
\bibitem [{\citenamefont {Kocsis}\ \emph {et~al.}(2018)\citenamefont {Kocsis},
  \citenamefont {Suyama}, \citenamefont {Tanaka},\ and\ \citenamefont
  {Yokoyama}}]{Kocsis:2017yty}%
  \BibitemOpen
  \bibfield  {author} {\bibinfo {author} {\bibfnamefont {B.}~\bibnamefont
  {Kocsis}}, \bibinfo {author} {\bibfnamefont {T.}~\bibnamefont {Suyama}},
  \bibinfo {author} {\bibfnamefont {T.}~\bibnamefont {Tanaka}}, \ and\ \bibinfo
  {author} {\bibfnamefont {S.}~\bibnamefont {Yokoyama}},\ }\href {\doibase
  10.3847/1538-4357/aaa7f4} {\bibfield  {journal} {\bibinfo  {journal}
  {Astrophys. J.}\ }\textbf {\bibinfo {volume} {854}},\ \bibinfo {pages} {41}
  (\bibinfo {year} {2018})},\ \Eprint {http://arxiv.org/abs/1709.09007}
  {arXiv:1709.09007 [astro-ph.CO]} \BibitemShut {NoStop}%
\bibitem [{\citenamefont {Ali-Haïmoud}\ \emph {et~al.}(2017)\citenamefont
  {Ali-Haïmoud}, \citenamefont {Kovetz},\ and\ \citenamefont
  {Kamionkowski}}]{Ali-Haimoud:2017rtz}%
  \BibitemOpen
  \bibfield  {author} {\bibinfo {author} {\bibfnamefont {Y.}~\bibnamefont
  {Ali-Haïmoud}}, \bibinfo {author} {\bibfnamefont {E.~D.}\ \bibnamefont
  {Kovetz}}, \ and\ \bibinfo {author} {\bibfnamefont {M.}~\bibnamefont
  {Kamionkowski}},\ }\href {\doibase 10.1103/PhysRevD.96.123523} {\bibfield
  {journal} {\bibinfo  {journal} {Phys. Rev.}\ }\textbf {\bibinfo {volume}
  {D96}},\ \bibinfo {pages} {123523} (\bibinfo {year} {2017})},\ \Eprint
  {http://arxiv.org/abs/1709.06576} {arXiv:1709.06576 [astro-ph.CO]}
  \BibitemShut {NoStop}%
\bibitem [{\citenamefont {Chen}\ and\ \citenamefont
  {Huang}(2018)}]{Chen:2018czv}%
  \BibitemOpen
  \bibfield  {author} {\bibinfo {author} {\bibfnamefont {Z.-C.}\ \bibnamefont
  {Chen}}\ and\ \bibinfo {author} {\bibfnamefont {Q.-G.}\ \bibnamefont
  {Huang}},\ }\href {\doibase 10.3847/1538-4357/aad6e2} {\bibfield  {journal}
  {\bibinfo  {journal} {Astrophys. J.}\ }\textbf {\bibinfo {volume} {864}},\
  \bibinfo {pages} {61} (\bibinfo {year} {2018})},\ \Eprint
  {http://arxiv.org/abs/1801.10327} {arXiv:1801.10327 [astro-ph.CO]}
  \BibitemShut {NoStop}%
\bibitem [{\citenamefont {Ballesteros}\ \emph {et~al.}(2018)\citenamefont
  {Ballesteros}, \citenamefont {Serpico},\ and\ \citenamefont
  {Taoso}}]{Ballesteros:2018swv}%
  \BibitemOpen
  \bibfield  {author} {\bibinfo {author} {\bibfnamefont {G.}~\bibnamefont
  {Ballesteros}}, \bibinfo {author} {\bibfnamefont {P.~D.}\ \bibnamefont
  {Serpico}}, \ and\ \bibinfo {author} {\bibfnamefont {M.}~\bibnamefont
  {Taoso}},\ }\href {\doibase 10.1088/1475-7516/2018/10/043} {\bibfield
  {journal} {\bibinfo  {journal} {JCAP}\ }\textbf {\bibinfo {volume} {1810}},\
  \bibinfo {pages} {043} (\bibinfo {year} {2018})},\ \Eprint
  {http://arxiv.org/abs/1807.02084} {arXiv:1807.02084 [astro-ph.CO]}
  \BibitemShut {NoStop}%
\bibitem [{\citenamefont {Raidal}\ \emph {et~al.}(2019)\citenamefont {Raidal},
  \citenamefont {Spethmann}, \citenamefont {Vaskonen},\ and\ \citenamefont
  {Veermäe}}]{Raidal:2018bbj}%
  \BibitemOpen
  \bibfield  {author} {\bibinfo {author} {\bibfnamefont {M.}~\bibnamefont
  {Raidal}}, \bibinfo {author} {\bibfnamefont {C.}~\bibnamefont {Spethmann}},
  \bibinfo {author} {\bibfnamefont {V.}~\bibnamefont {Vaskonen}}, \ and\
  \bibinfo {author} {\bibfnamefont {H.}~\bibnamefont {Veermäe}},\ }\href
  {\doibase 10.1088/1475-7516/2019/02/018} {\bibfield  {journal} {\bibinfo
  {journal} {JCAP}\ }\textbf {\bibinfo {volume} {1902}},\ \bibinfo {pages}
  {018} (\bibinfo {year} {2019})},\ \Eprint {http://arxiv.org/abs/1812.01930}
  {arXiv:1812.01930 [astro-ph.CO]} \BibitemShut {NoStop}%
\bibitem [{\citenamefont {Liu}\ \emph {et~al.}(2019{\natexlab{b}})\citenamefont
  {Liu}, \citenamefont {Guo},\ and\ \citenamefont {Cai}}]{Liu:2018ess}%
  \BibitemOpen
  \bibfield  {author} {\bibinfo {author} {\bibfnamefont {L.}~\bibnamefont
  {Liu}}, \bibinfo {author} {\bibfnamefont {Z.-K.}\ \bibnamefont {Guo}}, \ and\
  \bibinfo {author} {\bibfnamefont {R.-G.}\ \bibnamefont {Cai}},\ }\href
  {\doibase 10.1103/PhysRevD.99.063523} {\bibfield  {journal} {\bibinfo
  {journal} {Phys. Rev.}\ }\textbf {\bibinfo {volume} {D99}},\ \bibinfo {pages}
  {063523} (\bibinfo {year} {2019}{\natexlab{b}})},\ \Eprint
  {http://arxiv.org/abs/1812.05376} {arXiv:1812.05376 [astro-ph.CO]}
  \BibitemShut {NoStop}%
\bibitem [{\citenamefont {Liu}\ \emph {et~al.}(2019{\natexlab{c}})\citenamefont
  {Liu}, \citenamefont {Guo},\ and\ \citenamefont {Cai}}]{Liu:2019rnx}%
  \BibitemOpen
  \bibfield  {author} {\bibinfo {author} {\bibfnamefont {L.}~\bibnamefont
  {Liu}}, \bibinfo {author} {\bibfnamefont {Z.-K.}\ \bibnamefont {Guo}}, \ and\
  \bibinfo {author} {\bibfnamefont {R.-G.}\ \bibnamefont {Cai}},\ }\href
  {\doibase 10.1140/epjc/s10052-019-7227-0} {\bibfield  {journal} {\bibinfo
  {journal} {Eur. Phys. J.}\ }\textbf {\bibinfo {volume} {C79}},\ \bibinfo
  {pages} {717} (\bibinfo {year} {2019}{\natexlab{c}})},\ \Eprint
  {http://arxiv.org/abs/1901.07672} {arXiv:1901.07672 [astro-ph.CO]}
  \BibitemShut {NoStop}%
\bibitem [{\citenamefont {Young}\ and\ \citenamefont
  {Byrnes}(2019)}]{Young:2019gfc}%
  \BibitemOpen
  \bibfield  {author} {\bibinfo {author} {\bibfnamefont {S.}~\bibnamefont
  {Young}}\ and\ \bibinfo {author} {\bibfnamefont {C.~T.}\ \bibnamefont
  {Byrnes}},\ }\href@noop {} {\  (\bibinfo {year} {2019})},\ \Eprint
  {http://arxiv.org/abs/1910.06077} {arXiv:1910.06077 [astro-ph.CO]}
  \BibitemShut {NoStop}%
\bibitem [{\citenamefont {Vaskonen}\ and\ \citenamefont
  {Veermäe}(2019)}]{Vaskonen:2019jpv}%
  \BibitemOpen
  \bibfield  {author} {\bibinfo {author} {\bibfnamefont {V.}~\bibnamefont
  {Vaskonen}}\ and\ \bibinfo {author} {\bibfnamefont {H.}~\bibnamefont
  {Veermäe}},\ }\href@noop {} {\  (\bibinfo {year} {2019})},\ \Eprint
  {http://arxiv.org/abs/1908.09752} {arXiv:1908.09752 [astro-ph.CO]}
  \BibitemShut {NoStop}%
\bibitem [{\citenamefont {Garriga}\ and\ \citenamefont
  {Triantafyllou}(2019)}]{Garriga:2019vqu}%
  \BibitemOpen
  \bibfield  {author} {\bibinfo {author} {\bibfnamefont {J.}~\bibnamefont
  {Garriga}}\ and\ \bibinfo {author} {\bibfnamefont {N.}~\bibnamefont
  {Triantafyllou}},\ }\href {\doibase 10.1088/1475-7516/2019/09/043} {\bibfield
   {journal} {\bibinfo  {journal} {JCAP}\ }\textbf {\bibinfo {volume} {1909}},\
  \bibinfo {pages} {043} (\bibinfo {year} {2019})},\ \Eprint
  {http://arxiv.org/abs/1907.01455} {arXiv:1907.01455 [astro-ph.CO]}
  \BibitemShut {NoStop}%
\bibitem [{\citenamefont {O’Leary}\ \emph {et~al.}(2016)\citenamefont
  {O’Leary}, \citenamefont {Meiron},\ and\ \citenamefont
  {Kocsis}}]{O_Leary_2016}%
  \BibitemOpen
  \bibfield  {author} {\bibinfo {author} {\bibfnamefont {R.~M.}\ \bibnamefont
  {O’Leary}}, \bibinfo {author} {\bibfnamefont {Y.}~\bibnamefont {Meiron}}, \
  and\ \bibinfo {author} {\bibfnamefont {B.}~\bibnamefont {Kocsis}},\ }\href
  {\doibase 10.3847/2041-8205/824/1/l12} {\bibfield  {journal} {\bibinfo
  {journal} {The Astrophysical Journal}\ }\textbf {\bibinfo {volume} {824}},\
  \bibinfo {pages} {L12} (\bibinfo {year} {2016})}\BibitemShut {NoStop}%
\bibitem [{\citenamefont {Page}(2005)}]{Page:2004xp}%
  \BibitemOpen
  \bibfield  {author} {\bibinfo {author} {\bibfnamefont {D.~N.}\ \bibnamefont
  {Page}},\ }\href {\doibase 10.1088/1367-2630/7/1/203} {\bibfield  {journal}
  {\bibinfo  {journal} {New J. Phys.}\ }\textbf {\bibinfo {volume} {7}},\
  \bibinfo {pages} {203} (\bibinfo {year} {2005})},\ \Eprint
  {http://arxiv.org/abs/hep-th/0409024} {arXiv:hep-th/0409024 [hep-th]}
  \BibitemShut {NoStop}%
\bibitem [{\citenamefont {Chen}\ \emph
  {et~al.}(2018{\natexlab{b}})\citenamefont {Chen}, \citenamefont {Kim},
  \citenamefont {Sun},\ and\ \citenamefont {Tang}}]{Chen:2017mnm}%
  \BibitemOpen
  \bibfield  {author} {\bibinfo {author} {\bibfnamefont {C.-M.}\ \bibnamefont
  {Chen}}, \bibinfo {author} {\bibfnamefont {S.~P.}\ \bibnamefont {Kim}},
  \bibinfo {author} {\bibfnamefont {J.-R.}\ \bibnamefont {Sun}}, \ and\
  \bibinfo {author} {\bibfnamefont {F.-Y.}\ \bibnamefont {Tang}},\ }\href
  {\doibase 10.1016/j.physletb.2018.03.078} {\bibfield  {journal} {\bibinfo
  {journal} {Phys. Lett.}\ }\textbf {\bibinfo {volume} {B781}},\ \bibinfo
  {pages} {129} (\bibinfo {year} {2018}{\natexlab{b}})},\ \Eprint
  {http://arxiv.org/abs/1705.10629} {arXiv:1705.10629 [hep-th]} \BibitemShut
  {NoStop}%
\bibitem [{\citenamefont {Cai}\ and\ \citenamefont {Kim}(2014)}]{Cai:2014qba}%
  \BibitemOpen
  \bibfield  {author} {\bibinfo {author} {\bibfnamefont {R.-G.}\ \bibnamefont
  {Cai}}\ and\ \bibinfo {author} {\bibfnamefont {S.~P.}\ \bibnamefont {Kim}},\
  }\href {\doibase 10.1007/JHEP09(2014)072} {\bibfield  {journal} {\bibinfo
  {journal} {JHEP}\ }\textbf {\bibinfo {volume} {09}},\ \bibinfo {pages} {072}
  (\bibinfo {year} {2014})},\ \Eprint {http://arxiv.org/abs/1407.4569}
  {arXiv:1407.4569 [hep-th]} \BibitemShut {NoStop}%
\bibitem [{\citenamefont {Aharonov}\ \emph {et~al.}(1987)\citenamefont
  {Aharonov}, \citenamefont {Casher},\ and\ \citenamefont
  {Nussinov}}]{Aharonov:1987tp}%
  \BibitemOpen
  \bibfield  {author} {\bibinfo {author} {\bibfnamefont {Y.}~\bibnamefont
  {Aharonov}}, \bibinfo {author} {\bibfnamefont {A.}~\bibnamefont {Casher}}, \
  and\ \bibinfo {author} {\bibfnamefont {S.}~\bibnamefont {Nussinov}},\ }\href
  {\doibase 10.1016/0370-2693(87)91320-7} {\bibfield  {journal} {\bibinfo
  {journal} {Phys. Lett.}\ }\textbf {\bibinfo {volume} {B191}},\ \bibinfo
  {pages} {51} (\bibinfo {year} {1987})}\BibitemShut {NoStop}%
\bibitem [{\citenamefont {Pioline}\ and\ \citenamefont
  {Troost}(2005)}]{Pioline:2005pf}%
  \BibitemOpen
  \bibfield  {author} {\bibinfo {author} {\bibfnamefont {B.}~\bibnamefont
  {Pioline}}\ and\ \bibinfo {author} {\bibfnamefont {J.}~\bibnamefont
  {Troost}},\ }\href {\doibase 10.1088/1126-6708/2005/03/043} {\bibfield
  {journal} {\bibinfo  {journal} {JHEP}\ }\textbf {\bibinfo {volume} {03}},\
  \bibinfo {pages} {043} (\bibinfo {year} {2005})},\ \Eprint
  {http://arxiv.org/abs/hep-th/0501169} {arXiv:hep-th/0501169 [hep-th]}
  \BibitemShut {NoStop}%
\bibitem [{\citenamefont {Chen}\ \emph {et~al.}(2017)\citenamefont {Chen},
  \citenamefont {Kim}, \citenamefont {Sun},\ and\ \citenamefont
  {Tang}}]{Chen:2016caa}%
  \BibitemOpen
  \bibfield  {author} {\bibinfo {author} {\bibfnamefont {C.-M.}\ \bibnamefont
  {Chen}}, \bibinfo {author} {\bibfnamefont {S.~P.}\ \bibnamefont {Kim}},
  \bibinfo {author} {\bibfnamefont {J.-R.}\ \bibnamefont {Sun}}, \ and\
  \bibinfo {author} {\bibfnamefont {F.-Y.}\ \bibnamefont {Tang}},\ }\href
  {\doibase 10.1103/PhysRevD.95.044043} {\bibfield  {journal} {\bibinfo
  {journal} {Phys. Rev.}\ }\textbf {\bibinfo {volume} {D95}},\ \bibinfo {pages}
  {044043} (\bibinfo {year} {2017})},\ \Eprint
  {http://arxiv.org/abs/1607.02610} {arXiv:1607.02610 [hep-th]} \BibitemShut
  {NoStop}%
\bibitem [{\citenamefont {Bai}\ and\ \citenamefont
  {Orlofsky}(2020)}]{Bai:2019zcd}%
  \BibitemOpen
  \bibfield  {author} {\bibinfo {author} {\bibfnamefont {Y.}~\bibnamefont
  {Bai}}\ and\ \bibinfo {author} {\bibfnamefont {N.}~\bibnamefont {Orlofsky}},\
  }\href {\doibase 10.1103/PhysRevD.101.055006} {\bibfield  {journal} {\bibinfo
   {journal} {Phys. Rev. D}\ }\textbf {\bibinfo {volume} {101}},\ \bibinfo
  {pages} {055006} (\bibinfo {year} {2020})},\ \Eprint
  {http://arxiv.org/abs/1906.04858} {arXiv:1906.04858 [hep-ph]} \BibitemShut
  {NoStop}%
\bibitem [{\citenamefont {Aghanim}\ \emph {et~al.}(2018)\citenamefont {Aghanim}
  \emph {et~al.}}]{Aghanim:2018eyx}%
  \BibitemOpen
  \bibfield  {author} {\bibinfo {author} {\bibfnamefont {N.}~\bibnamefont
  {Aghanim}} \emph {et~al.} (\bibinfo {collaboration} {Planck}),\ }\href@noop
  {} {\  (\bibinfo {year} {2018})},\ \Eprint {http://arxiv.org/abs/1807.06209}
  {arXiv:1807.06209 [astro-ph.CO]} \BibitemShut {NoStop}%
\bibitem [{\citenamefont {Peters}\ and\ \citenamefont
  {Mathews}(1963)}]{Peters:1963ux}%
  \BibitemOpen
  \bibfield  {author} {\bibinfo {author} {\bibfnamefont {P.~C.}\ \bibnamefont
  {Peters}}\ and\ \bibinfo {author} {\bibfnamefont {J.}~\bibnamefont
  {Mathews}},\ }\href {\doibase 10.1103/PhysRev.131.435} {\bibfield  {journal}
  {\bibinfo  {journal} {Phys. Rev.}\ }\textbf {\bibinfo {volume} {131}},\
  \bibinfo {pages} {435} (\bibinfo {year} {1963})}\BibitemShut {NoStop}%
\bibitem [{\citenamefont {Peters}(1964)}]{Peters:1964zz}%
  \BibitemOpen
  \bibfield  {author} {\bibinfo {author} {\bibfnamefont {P.~C.}\ \bibnamefont
  {Peters}},\ }\href {\doibase 10.1103/PhysRev.136.B1224} {\bibfield  {journal}
  {\bibinfo  {journal} {Phys. Rev.}\ }\textbf {\bibinfo {volume} {136}},\
  \bibinfo {pages} {B1224} (\bibinfo {year} {1964})}\BibitemShut {NoStop}%
\bibitem [{\citenamefont {Hawking}(1975)}]{Hawking:1974sw}%
  \BibitemOpen
  \bibfield  {author} {\bibinfo {author} {\bibfnamefont {S.}~\bibnamefont
  {Hawking}},\ }\href {\doibase 10.1007/BF02345020} {\bibfield  {journal}
  {\bibinfo  {journal} {Commun. Math. Phys.}\ }\textbf {\bibinfo {volume}
  {43}},\ \bibinfo {pages} {199} (\bibinfo {year} {1975})},\ \bibinfo {note}
  {[Erratum: Commun.Math.Phys. 46, 206 (1976)]}\BibitemShut {NoStop}%
\bibitem [{\citenamefont {Page}(1976)}]{Page:1976ki}%
  \BibitemOpen
  \bibfield  {author} {\bibinfo {author} {\bibfnamefont {D.~N.}\ \bibnamefont
  {Page}},\ }\href {\doibase 10.1103/PhysRevD.14.3260} {\bibfield  {journal}
  {\bibinfo  {journal} {Phys. Rev.}\ }\textbf {\bibinfo {volume} {D14}},\
  \bibinfo {pages} {3260} (\bibinfo {year} {1976})}\BibitemShut {NoStop}%
\bibitem [{\citenamefont {Cardoso}\ \emph {et~al.}(2016)\citenamefont
  {Cardoso}, \citenamefont {Macedo}, \citenamefont {Pani},\ and\ \citenamefont
  {Ferrari}}]{Cardoso_2016}%
  \BibitemOpen
  \bibfield  {author} {\bibinfo {author} {\bibfnamefont {V.}~\bibnamefont
  {Cardoso}}, \bibinfo {author} {\bibfnamefont {C.~F.}\ \bibnamefont {Macedo}},
  \bibinfo {author} {\bibfnamefont {P.}~\bibnamefont {Pani}}, \ and\ \bibinfo
  {author} {\bibfnamefont {V.}~\bibnamefont {Ferrari}},\ }\href {\doibase
  10.1088/1475-7516/2016/05/054} {\bibfield  {journal} {\bibinfo  {journal}
  {Journal of Cosmology and Astroparticle Physics}\ }\textbf {\bibinfo {volume}
  {2016}},\ \bibinfo {pages} {054–054} (\bibinfo {year} {2016})}\BibitemShut
  {NoStop}%
\bibitem [{\citenamefont {Carr}\ \emph {et~al.}(2020)\citenamefont {Carr},
  \citenamefont {Kohri}, \citenamefont {Sendouda},\ and\ \citenamefont
  {Yokoyama}}]{Carr:2020gox}%
  \BibitemOpen
  \bibfield  {author} {\bibinfo {author} {\bibfnamefont {B.}~\bibnamefont
  {Carr}}, \bibinfo {author} {\bibfnamefont {K.}~\bibnamefont {Kohri}},
  \bibinfo {author} {\bibfnamefont {Y.}~\bibnamefont {Sendouda}}, \ and\
  \bibinfo {author} {\bibfnamefont {J.}~\bibnamefont {Yokoyama}},\ }\href@noop
  {} {\  (\bibinfo {year} {2020})},\ \Eprint {http://arxiv.org/abs/2002.12778}
  {arXiv:2002.12778 [astro-ph.CO]} \BibitemShut {NoStop}%
\bibitem [{\citenamefont {Desjacques}\ and\ \citenamefont
  {Riotto}(2018)}]{Desjacques:2018wuu}%
  \BibitemOpen
  \bibfield  {author} {\bibinfo {author} {\bibfnamefont {V.}~\bibnamefont
  {Desjacques}}\ and\ \bibinfo {author} {\bibfnamefont {A.}~\bibnamefont
  {Riotto}},\ }\href {\doibase 10.1103/PhysRevD.98.123533} {\bibfield
  {journal} {\bibinfo  {journal} {Phys. Rev.}\ }\textbf {\bibinfo {volume}
  {D98}},\ \bibinfo {pages} {123533} (\bibinfo {year} {2018})},\ \Eprint
  {http://arxiv.org/abs/1806.10414} {arXiv:1806.10414 [astro-ph.CO]}
  \BibitemShut {NoStop}%
\bibitem [{\citenamefont {Bringmann}\ \emph {et~al.}(2019)\citenamefont
  {Bringmann}, \citenamefont {Depta}, \citenamefont {Domcke},\ and\
  \citenamefont {Schmidt-Hoberg}}]{Bringmann:2018mxj}%
  \BibitemOpen
  \bibfield  {author} {\bibinfo {author} {\bibfnamefont {T.}~\bibnamefont
  {Bringmann}}, \bibinfo {author} {\bibfnamefont {P.~F.}\ \bibnamefont
  {Depta}}, \bibinfo {author} {\bibfnamefont {V.}~\bibnamefont {Domcke}}, \
  and\ \bibinfo {author} {\bibfnamefont {K.}~\bibnamefont {Schmidt-Hoberg}},\
  }\href {\doibase 10.1103/PhysRevD.99.063532} {\bibfield  {journal} {\bibinfo
  {journal} {Phys. Rev.}\ }\textbf {\bibinfo {volume} {D99}},\ \bibinfo {pages}
  {063532} (\bibinfo {year} {2019})},\ \Eprint
  {http://arxiv.org/abs/1808.05910} {arXiv:1808.05910 [astro-ph.CO]}
  \BibitemShut {NoStop}%
\bibitem [{\citenamefont {Suyama}\ and\ \citenamefont
  {Yokoyama}(2019)}]{Suyama:2019cst}%
  \BibitemOpen
  \bibfield  {author} {\bibinfo {author} {\bibfnamefont {T.}~\bibnamefont
  {Suyama}}\ and\ \bibinfo {author} {\bibfnamefont {S.}~\bibnamefont
  {Yokoyama}},\ }\href {\doibase 10.1093/ptep/ptz105} {\bibfield  {journal}
  {\bibinfo  {journal} {PTEP}\ }\textbf {\bibinfo {volume} {2019}},\ \bibinfo
  {pages} {103E02} (\bibinfo {year} {2019})},\ \Eprint
  {http://arxiv.org/abs/1906.04958} {arXiv:1906.04958 [astro-ph.CO]}
  \BibitemShut {NoStop}%
\bibitem [{\citenamefont {Olson}\ and\ \citenamefont
  {Bailyn}(1975)}]{Olson:1975ts}%
  \BibitemOpen
  \bibfield  {author} {\bibinfo {author} {\bibfnamefont {E.}~\bibnamefont
  {Olson}}\ and\ \bibinfo {author} {\bibfnamefont {M.}~\bibnamefont {Bailyn}},\
  }\href {\doibase 10.1103/PhysRevD.12.3030} {\bibfield  {journal} {\bibinfo
  {journal} {Phys. Rev. D}\ }\textbf {\bibinfo {volume} {12}},\ \bibinfo
  {pages} {3030} (\bibinfo {year} {1975})}\BibitemShut {NoStop}%
\bibitem [{\citenamefont {{Bally}}\ and\ \citenamefont
  {{Harrison}}(1978)}]{1978ApJ...220..743B}%
  \BibitemOpen
  \bibfield  {author} {\bibinfo {author} {\bibfnamefont {J.}~\bibnamefont
  {{Bally}}}\ and\ \bibinfo {author} {\bibfnamefont {E.~R.}\ \bibnamefont
  {{Harrison}}},\ }\href {\doibase 10.1086/155961} {\bibfield  {journal}
  {\bibinfo  {journal} {\apj}\ }\textbf {\bibinfo {volume} {220}},\ \bibinfo
  {pages} {743} (\bibinfo {year} {1978})}\BibitemShut {NoStop}%
\bibitem [{\citenamefont {Ray}\ \emph {et~al.}(2003)\citenamefont {Ray},
  \citenamefont {Espindola}, \citenamefont {Malheiro}, \citenamefont {Lemos},\
  and\ \citenamefont {Zanchin}}]{Ray:2003gt}%
  \BibitemOpen
  \bibfield  {author} {\bibinfo {author} {\bibfnamefont {S.}~\bibnamefont
  {Ray}}, \bibinfo {author} {\bibfnamefont {A.~L.}\ \bibnamefont {Espindola}},
  \bibinfo {author} {\bibfnamefont {M.}~\bibnamefont {Malheiro}}, \bibinfo
  {author} {\bibfnamefont {J.~P.}\ \bibnamefont {Lemos}}, \ and\ \bibinfo
  {author} {\bibfnamefont {V.~T.}\ \bibnamefont {Zanchin}},\ }\href {\doibase
  10.1103/PhysRevD.68.084004} {\bibfield  {journal} {\bibinfo  {journal} {Phys.
  Rev. D}\ }\textbf {\bibinfo {volume} {68}},\ \bibinfo {pages} {084004}
  (\bibinfo {year} {2003})},\ \Eprint {http://arxiv.org/abs/astro-ph/0307262}
  {arXiv:astro-ph/0307262} \BibitemShut {NoStop}%
\bibitem [{\citenamefont {Zhang}(2016)}]{Zhang:2016rli}%
  \BibitemOpen
  \bibfield  {author} {\bibinfo {author} {\bibfnamefont {B.}~\bibnamefont
  {Zhang}},\ }\href {\doibase 10.3847/2041-8205/827/2/L31} {\bibfield
  {journal} {\bibinfo  {journal} {Astrophys. J.}\ }\textbf {\bibinfo {volume}
  {827}},\ \bibinfo {pages} {L31} (\bibinfo {year} {2016})},\ \Eprint
  {http://arxiv.org/abs/1602.04542} {arXiv:1602.04542 [astro-ph.HE]}
  \BibitemShut {NoStop}%
\bibitem [{\citenamefont {Zaja\v~cek}\ and\ \citenamefont
  {Tursunov}(2019)}]{Zajacek:2019kla}%
  \BibitemOpen
  \bibfield  {author} {\bibinfo {author} {\bibfnamefont {M.}~\bibnamefont
  {Zaja\v~cek}}\ and\ \bibinfo {author} {\bibfnamefont {A.}~\bibnamefont
  {Tursunov}},\ }\href@noop {} {\  (\bibinfo {year} {2019})},\ \Eprint
  {http://arxiv.org/abs/1904.04654} {arXiv:1904.04654 [astro-ph.GA]}
  \BibitemShut {NoStop}%
\bibitem [{\citenamefont {Gibbons}(1977)}]{Gibbons:1976sm}%
  \BibitemOpen
  \bibfield  {author} {\bibinfo {author} {\bibfnamefont {G.~W.}\ \bibnamefont
  {Gibbons}},\ }\href {\doibase 10.1103/PhysRevD.15.3530} {\bibfield  {journal}
  {\bibinfo  {journal} {Phys. Rev.}\ }\textbf {\bibinfo {volume} {D15}},\
  \bibinfo {pages} {3530} (\bibinfo {year} {1977})}\BibitemShut {NoStop}%
\bibitem [{\citenamefont {Kim}(2019)}]{Kim:2019joy}%
  \BibitemOpen
  \bibfield  {author} {\bibinfo {author} {\bibfnamefont {S.~P.}\ \bibnamefont
  {Kim}},\ }in\ \href@noop {} {\emph {\bibinfo {booktitle} {{15th Marcel
  Grossmann Meeting on Recent Developments in Theoretical and Experimental
  General Relativity, Astrophysics, and Relativistic Field Theories (MG15)
  Rome, Italy, July 1-7, 2018}}}}\ (\bibinfo {year} {2019})\ \Eprint
  {http://arxiv.org/abs/1905.13439} {arXiv:1905.13439 [gr-qc]} \BibitemShut
  {NoStop}%
\bibitem [{\citenamefont {Romans}(1992)}]{Romans:1991nq}%
  \BibitemOpen
  \bibfield  {author} {\bibinfo {author} {\bibfnamefont {L.~J.}\ \bibnamefont
  {Romans}},\ }\href {\doibase 10.1016/0550-3213(92)90684-4} {\bibfield
  {journal} {\bibinfo  {journal} {Nucl. Phys.}\ }\textbf {\bibinfo {volume}
  {B383}},\ \bibinfo {pages} {395} (\bibinfo {year} {1992})},\ \Eprint
  {http://arxiv.org/abs/hep-th/9203018} {arXiv:hep-th/9203018 [hep-th]}
  \BibitemShut {NoStop}%
\bibitem [{\citenamefont {Cai}\ \emph {et~al.}(1998)\citenamefont {Cai},
  \citenamefont {Ji},\ and\ \citenamefont {Soh}}]{Cai:1997ih}%
  \BibitemOpen
  \bibfield  {author} {\bibinfo {author} {\bibfnamefont {R.-G.}\ \bibnamefont
  {Cai}}, \bibinfo {author} {\bibfnamefont {J.-Y.}\ \bibnamefont {Ji}}, \ and\
  \bibinfo {author} {\bibfnamefont {K.-S.}\ \bibnamefont {Soh}},\ }\bibfield
  {booktitle} {\emph {\bibinfo {booktitle} {{Topology of the universe.
  Proceedings, Conference, Cleveland, USA, October 17-19, 1997}}},\ }\href
  {\doibase 10.1088/0264-9381/15/9/023} {\bibfield  {journal} {\bibinfo
  {journal} {Class. Quant. Grav.}\ }\textbf {\bibinfo {volume} {15}},\ \bibinfo
  {pages} {2783} (\bibinfo {year} {1998})},\ \Eprint
  {http://arxiv.org/abs/gr-qc/9708062} {arXiv:gr-qc/9708062 [gr-qc]}
  \BibitemShut {NoStop}%
\bibitem [{\citenamefont {Montero}\ \emph {et~al.}(2019)\citenamefont
  {Montero}, \citenamefont {Van~Riet},\ and\ \citenamefont
  {Venken}}]{Montero:2019ekk}%
  \BibitemOpen
  \bibfield  {author} {\bibinfo {author} {\bibfnamefont {M.}~\bibnamefont
  {Montero}}, \bibinfo {author} {\bibfnamefont {T.}~\bibnamefont {Van~Riet}}, \
  and\ \bibinfo {author} {\bibfnamefont {G.}~\bibnamefont {Venken}},\
  }\href@noop {} {\  (\bibinfo {year} {2019})},\ \Eprint
  {http://arxiv.org/abs/1910.01648} {arXiv:1910.01648 [hep-th]} \BibitemShut
  {NoStop}%
\bibitem [{\citenamefont {Maleknejad}\ \emph {et~al.}(2013)\citenamefont
  {Maleknejad}, \citenamefont {Sheikh-Jabbari},\ and\ \citenamefont
  {Soda}}]{Maleknejad:2012fw}%
  \BibitemOpen
  \bibfield  {author} {\bibinfo {author} {\bibfnamefont {A.}~\bibnamefont
  {Maleknejad}}, \bibinfo {author} {\bibfnamefont {M.~M.}\ \bibnamefont
  {Sheikh-Jabbari}}, \ and\ \bibinfo {author} {\bibfnamefont {J.}~\bibnamefont
  {Soda}},\ }\href {\doibase 10.1016/j.physrep.2013.03.003} {\bibfield
  {journal} {\bibinfo  {journal} {Phys. Rept.}\ }\textbf {\bibinfo {volume}
  {528}},\ \bibinfo {pages} {161} (\bibinfo {year} {2013})},\ \Eprint
  {http://arxiv.org/abs/1212.2921} {arXiv:1212.2921 [hep-th]} \BibitemShut
  {NoStop}%
\end{thebibliography}%

\end{document}